\PassOptionsToPackage{hyphens}{url}  

\documentclass[fleqn,usenatbib]{mnras}

\usepackage{newtxtext,newtxmath}
\usepackage[T1]{fontenc}

\DeclareRobustCommand{\VAN}[3]{#2}
\let\VANthebibliography\thebibliography
\def\thebibliography{\DeclareRobustCommand{\VAN}[3]{##3}\VANthebibliography}

\usepackage{amsmath}
\usepackage{bm}
\usepackage{booktabs}
\usepackage{cleveref}
\usepackage{duckuments}  
\usepackage[acronym]{glossaries}
\usepackage{graphicx}
\usepackage{hyperref}
\usepackage{multirow}
\usepackage{physics}
\usepackage{siunitx}
\usepackage[normalem]{ulem}  

\newacronym{abc}{ABC}{approximate Bayesian computation}
\newacronym{cnn}{CNN}{convolutional neural network}
\newacronym{dm}{DM}{dark matter}
\newacronym{gan}{GAN}{generative adversarial network}
\newacronym{gr}{GR}{general relativity}
\newacronym{hmc}{HMC}{Hamiltonian Monte Carlo}
\newacronym{hpd}{HPD}{highest posterior density}
\newacronym{los}{LOS}{line-of-sight}
\newacronym{mcmc}{MCMC}{Markov-chain Monte Carlo}
\newacronym{mlp}{MLP}{multilayer perceptron}
\newacronym{nfw}{NFW}{Navarro-Frenk-White}
\newacronym{nre}{NRE}{neural ratio estimation}
\newacronym{psf}{PSF}{point-spread function}
\newacronym{sbi}{SBI}{simulation-based inference}
\newacronym{shmf}{SHMF}{subhalo mass function}
\newacronym{sie}{SIE}{singular isothermal ellipsoid}
\newacronym{sple}{SPLE}{singular power law ellipsoid}
\newacronym{tmnre}{TMNRE}{truncated marginal neural ratio estimation}
\newacronym{vit}{ViT}{Vision Transformer}

\newcommand{\data}{\ensuremath{\bm{x}}}
\newcommand{\hypgeom}{\ensuremath{{}_2\mathrm{F}_1}}
\newcommand{\msun}{\ensuremath{\mathrm{M}_\odot}}
\newcommand{\param}{\ensuremath{\bm{\theta}}}
\newcommand{\plens}{\ensuremath{\bm{\eta}_\mathrm{lens}}}
\newcommand{\psrc}{\ensuremath{\bm{\eta}_\mathrm{src}}}
\newcommand{\psub}{\ensuremath{\bm{\vartheta}_\mathrm{sub}}}

\DeclareMathOperator{\uniform}{\mathcal{U}}

\DeclareSIUnit\solarmass{\msun}

\title[The effect of perturbers on subhalo measurements]{One never walks alone: the effect of the perturber population on subhalo measurements in strong gravitational lenses}

\author[A. Coogan et al.]{Adam Coogan$^{1,2,3}$\thanks{adam.coogan@umontreal.ca},
Noemi Anau Montel$^{3}$\thanks{n.anaumontel@uva.nl},
Konstantin Karchev$^{3,4}$,
Meiert W. Grootes$^{5}$,
\newauthor
Francesco Nattino$^{5}$,
Christoph Weniger$^{3}$\thanks{c.weniger@uva.nl}
\\
$^{1}$Département de Physique, Université de Montréal, 1375 Avenue Thérèse-Lavoie-Roux, Montréal, QC H2V 0B3, Canada\\
$^{2}$Mila – Quebec AI Institute, 6666 St-Urbain, 200, Montreal, QC, H2S 3H1\\
$^{3}$GRAPPA (Gravitation Astroparticle Physics Amsterdam), University of Amsterdam, Science Park 904, 1098 XH Amsterdam, The Netherlands\\
$^{4}$SISSA (Scuola Internazionale Superiore di Studi Avanzati), via Bonomea 265, I-34136 Trieste, Italy\\
$^{5}$Netherlands eScience Center, Science Park 402 (Matrix III), 1098 XH Amsterdam, The Netherlands
}

\date{Accepted XXX. Received YYY; in original form ZZZ}

\pubyear{2021}

\begin{document}
\label{firstpage}
\pagerange{\pageref{firstpage}--\pageref{lastpage}}
\maketitle

\begin{abstract}
    Analyses of extended arcs in strong gravitational lensing images to date have constrained the properties of dark matter by measuring the parameters of one or two individual subhalos. However, since such analyses are reliant on likelihood-based methods like Markov-chain Monte Carlo or nested sampling, they require various compromises to the realism of lensing models for the sake of computational tractability, such as ignoring the numerous other subhalos and line-of-sight halos in the system, assuming a particular form for the source model and requiring the noise to have a known likelihood function. Here we show that a simulation-based inference method called \gls*{tmnre} makes it possible to relax these requirements by training neural networks to directly compute marginal posteriors for subhalo parameters from lensing images. By performing a set of inference tasks on mock data, we verify the accuracy of \gls*{tmnre} and show it can compute posteriors for subhalo parameters marginalized over populations of hundreds of subhalos and line-of-sight halos, as well as lens and source uncertainties. We also find the MLP Mixer network works far better for such tasks than the convolutional architectures explored in other lensing analyses. Furthermore, we show that since \gls*{tmnre} learns a posterior function it enables direct statistical checks that would be extremely expensive with likelihood-based methods. Our results show that \gls*{tmnre} is well-suited for analyzing complex lensing data, and that the full subhalo and line-of-sight halo population must be included when measuring the properties of individual dark matter substructures.
\end{abstract}


\section{Introduction}

Determining the microphysical properties of the \gls*{dm} comprising about $85\%$ of the universe's mass is one of the key problems in physics. The distribution of \gls*{dm} on scales larger than dwarf galaxies is well-characterized and consistent with \gls*{dm} behaving as an approximately cold, collisionless, classical fluid (see e.g. \citet{Profumo:2017hqp} for an overview). On the other hand, the distribution of \gls*{dm} on smaller scales is currently only roughly mapped out. At present, there is continued debate over whether the known abundance of dwarf galaxies and the density profiles of low-mass galaxies are in tension with the predictions of $\Lambda$CDM (respectively dubbed the missing satellites problem \citep{Moore:1999nt,Klypin:1999uc} and the cusp-core problem \citep{deBlok:1997zlw}, reviewed in \citet{Bullock:2017xww}). \Gls*{dm} models which are warm instead of cold \citep{Colin:2000dn,Hogan:2000bv}, collisional instead of collisionless \citep{Spergel:1999mh}, or quantum on macroscopic scales rather than classical \citep{Hu:2000ke} predict a diverse array of possible configurations of low-mass halos and could potentially resolve these tensions \citep{Buckley:2017ijx}.

Unfortunately, light \gls*{dm} halos are difficult to probe as they are not expected to accumulate enough baryonic matter to form stars \citep{Efstathiou:1992zz,Fitts:2016usl}. If \gls*{dm} has significant self-interactions, such halos might be detectable by searching for the self-annihilation or decay products of \gls*{dm}. However, even if such interactions are not present, light halos can potentially be probed through their irreducible gravitational effects. In this work we study one such probe: galaxy-galaxy strong gravitational lensing.

In galaxy-galaxy strong lenses the light from a \emph{source} galaxy is dramatically distorted into a ring shape by the mass of a \emph{lens} galaxy lying close to the line of sight to the source. This leads to multiple magnified and distorted images of the source, as explained by general relativity. A \emph{perturber} (i.e., a subhalo or \gls*{los} halo lying somewhere between the observer and source) positioned near one of these images contributes additional, much more localized distortions. By carefully analyzing the relationship between the multiple images of the source, the distortions from a perturber can be disentangled from possible variations in the source light and its properties can be measured. From measuring the distributions of perturbers' masses and other parameters, it is possible to infer population-level properties like the (sub)halo mass function parameters, which are dictated by the fundamental properties of dark matter.

Such analyses have been developed for two types of lensing systems: quadruply-lensed quasars (``quads'') and lenses with extended arcs. In the former, the source is a nearly point-like quasar that is lensed into four compact images. These images' positions and flux ratios comprise the summary statistics for these systems. The presence of a perturber near one of these images would cause anomalies in the ratios of their fluxes relative to what would be predicted assuming a smooth lens mass distribution. Evidence for flux ratio anomalies due to perturber was first found in \citet{Mao:1997ek}; \citet{Dalal:2001fq} developed the first statistical analysis to measure perturbers' properties from flux ratios.

Here we focus on \emph{gravitational imaging}, which refers to the analysis of lenses with extended arcs \citep{Koopmans:2005ig,Vegetti:2008eg,2009MNRAS.400.1583V}. The observation in this case consists of a whole image. On one hand, such images cover a larger area of the sky than the four point-like images in quads, potentially providing more sensitivity to detect perturbations due to perturber. On the other hand, extracting this information requires modeling the source galaxy's light, which generally has a complex morphology. 
Gravitational imaging has so far yielded several detections of $\sim\SI{e9}{\solarmass}$ perturbers using deep, high-resolution observations in the optical from the Hubble Space Telescope and Keck as well as in radio data from Atacama Large Millimeter/submillimeter Array \citep{2010MNRAS.407..225V,2010MNRAS.408.1969V,2012Natur.481..341V,Hezaveh:2016ltk}. Near-future telescopes such as the Rubin Observatory, Euclid, JWST and the Extremely Large Telescope will greatly increase the quality of data suitable for gravitational imaging analyses as well as its quantity (from $\order{100}$ to $\order{10^5}$ images \citep{Collett:2015roa}).

Established gravitational imaging analyses such as the method in \citet{Vegetti:2008eg,Hezaveh:2016ltk} use \emph{likelihood-based} inference to infer the properties of perturbers. Measurements and non-detections of individual perturbers can be converted to constraints on the (sub)halo mass function and thus dark matter's properties. The central mathematical object in such approaches is the likelihood, a probabilistic model $p(\data | \param)$ for the data $\data$ given some parameters $\param = (\plens, \psrc, \psub, \param_\mathrm{other})$ for the lens, source, perturber and possibly other (hyper)parameters.\footnote{
    For example, the hyperparameters could include the pixel size for pixelated sources or strength of source regularization.
} Statistical inference of perturber parameters $\psub$ such as mass and position given an observation $\data_0$ amounts to computing marginal posteriors $p(\psub | \data_0)$ by means of \gls*{mcmc} or nested sampling~\citep{2004AIPC..735..395S}. Likelihood-based inference tools do not directly produce marginal posteriors but instead compute the \emph{joint posterior} $p(\param | \data_0)$, which must then be marginalized over.

The computational expense of sampling from the joint posterior imposes restrictions on the realism of lensing models that can be analyzed. One such restriction common to most analyses is to assume no more than two perturbers are present in each image. Allowing for $n$ perturbers would cause the joint posterior to become highly multimodal, with approximately $n!$ modes due exact invariance of the observation under relabeling of perturbers. Transdimensional \gls*{mcmc} methods provide an inroad into this problem by inferring the probabilities of different possible populations of perturbers \citep{Brewer:2015yya,Daylan:2017kfh}, albeit at substantial computational cost. Another approach is to circumvent measuring individual perturbers by instead engineering summary statistics such as the power spectrum of the residuals between the image and best-fit reconstruction excluding substructure and relating them to the (sub)halo mass function parameters \citep{Hezaveh:2014aoa,Bayer:2018vhy,DiazRivero:2017xkd,CaganSengul:2020nat}. It is unknown how much information such approaches discard, and more generally unknown how large an impact ignoring all but one perturber has on measurements.

Likelihood-based analyses also typically assume a particular form of the noise and source model so that the source uncertainties can be excluded from the sampling and marginalized over analytically~\citep{Hezaveh:2016ltk,Vegetti:2008eg,2010MNRAS.408.1969V,2010MNRAS.407..225V,2012Natur.481..341V}. This makes it difficult to explore more complex source models described by e.g.\@ generative machine learning methods or noise artifacts like cosmic ray streaks that cannot be described by an analytic likelihood.

An additional difficulty with likelihood-based analyses is that each run of \gls*{mcmc} or nested sampling produces posterior samples for just a single observation. Directly exploring the systematics, biases and other statistical properties of a particular lensing model is thus extremely time-consuming, necessitating rerunning posterior sampling many times for different input observations. This also makes analyses such as mapping perturber measurement sensitivity costly.
\newline

\emph{In this work}, we demonstrate that a \gls*{sbi} \citep{2020PNAS..11730055C} method called \glsfirst*{tmnre} \citep{tmnre,swyft} can circumvent these inference challenges to measure the properties of individual perturbers. \gls*{sbi} refers to a class of statistical inference methods that use the output of a stochastic simulator that need not have a known likelihood. \Gls*{nre} \citep{aalr} in particular trains a neural network to map from observations directly to \emph{marginal posteriors} for a specified subset of model parameters (e.g. the position and mass of a perturber). This bypasses the requirement of likelihood-based inference to sample the joint posterior. In contrast to methods like \gls*{abc}, this also removes the need to engineer summary statistics~\citep{2020arXiv201013221H} as they are in effect learned directly from the training data. Since \gls*{nre} learns a marginal posterior \emph{function}, it is straightforward to check the statistical properties of the inference results for different observations. \gls*{tmnre} further extends \gls*{nre} by focusing training data generation in the regions of parameter space most relevant for analyzing a particular observation over a sequence of inference rounds. This substantially reduces the number of simulations required to train the inference network as well as the required network complexity.

Several other works have applied machine learning and \gls*{sbi} to substructure lensing. We recently demonstrated that \gls*{tmnre} can measure the cutoff in the warm \gls*{dm} \gls*{shmf} directly from images by combining multiple observations generated with a simple simulator \citep{Montel:2022fhv}. In \citet{Zhang:2022djp} a likelihood-ratio estimation technique similar to \gls*{tmnre} was employed to measure density profile parameters of subhalos from images. \citet{Wagner-Carena:2022mrn} recently applied neural posterior estimation to measure the \gls*{shmf} normalization in mock lensing images using real galaxy images as sources. \citet{Brehmer:2019jyt} utilized a ``likelihood-based'' \gls*{sbi} method requiring the simulator's score\footnote{
    The score is the derivative of the log-likelihood for a given observation with respect to the model's parameters.
} to measure the slope and normalization of a \gls*{shmf} in simple mock images. In \citet{Ostdiek:2020cqz,Ostdiek:2020mvo} image segmentation was used to classify whether each pixel in an image contained a subhalo in a given mass bin. Classifiers were also used in \citet{Alexander:2019puy} to distinguish between different \gls*{dm} models based on their lensing signatures.

The present work on measuring individual perturbers complements these efforts in several ways. First, it offers a path towards cross-checking current substructure measurements under different modeling assumptions. Second, inference based on perturbers provides a level of interpretability beyond measuring \gls*{shmf} parameters directly from images, and moreover the opportunity to test different \gls*{dm} models through measuring the properties of individual subhalos. Third, measuring the heaviest subhalos in an observation enables modeling them explicitly in lensing simulations, which could reduce the training data requirements and improve inference accuracy for direct \gls*{shmf} measurements.

This paper is organized as follows. In \cref{sec:model} we explain our lensing model, which uses an analytic source and main lens in conjunction with well-motivated perturber models. In \cref{sec:inference} we review \gls*{tmnre}. Our analysis begins in \cref{sec:sp}, where we show that \gls*{tmnre} is capable of recovering posteriors for a subhalo's mass and position in the limit where they are analytically-calculable. In the other analyses in \cref{sec:results} we gradually complexify our inference tasks, first accounting for the fact that the source and lens parameters are unknown and later by incorporating a population of light perturbers to marginalize over. This work will help form the basis for \gls*{tmnre}-based measurement of light \gls*{dm} halos in existing and future lensing data.

\section{Modeling strong lens observations}
\label{sec:model}

Here we summarize the source, main lens, perturbers and instrument models we use to simulate mock images of gravitational lenses. We implement our lensing model in \texttt{PyTorch} \citep{torch} so that we can leverage GPUs to rapidly generate large numbers of observations.

Before delving into modeling details we briefly summarize the key points of the physics of gravitational lensing, referring the reader to e.g.~\citet{lensing} for a more detailed overview. We assume that mass densities are low enough to treat the gravitational field of the matter in the image plane in the Newtonian approximation of \gls*{gr}. In this case the metric is fully characterized by the lens' gravitational potential $\psi$. We also adopt the thin lens approximation, which assumes all the lens mass lies in a single \emph{image plane} and all the source light is emitted from a \emph{source plane}. We use $\bm{\xi}$ and $\bm{x}$ as two-dimensional angular coordinates in the image and source planes respectively and use $z$ to indicate distances along the orthogonal dimension. Since the image plane covers a small angular patch of the sky and the lensing deflections are small in the Newtonian limit, the coordinate system can be treated as Cartesian.

In this setting, the lens' matter distribution can be described by its surface density
\begin{equation}
    \Sigma(\bm{\xi}) = \int \dd{z} \rho(\bm{\xi}, z) \, ,
\end{equation}
where $\rho$ is the lens' three-dimensional mass density and $z$ is its redshift. The source-plane coordinate to which a light ray through the image plane traces back is given by the \emph{lens equation}
\begin{equation}
    \bm{x} = \bm{\xi} - \bm{\alpha}(\bm{\xi}) \, .
\end{equation}
Here $\bm{\alpha}$ is the \emph{deflection field} of the lens, which can be computed through the integral
\begin{equation}
    \bm{\alpha}(\bm{\xi}) = \frac{4 G}{c^2} \frac{D_{LS}}{D_L \, D_S} \int \dd[2]{(D_L \bm{\xi}')} \frac{\bm{\xi} - \bm{\xi}'}{|\bm{\xi} - \bm{\xi}'|^2} \Sigma(\bm{\xi}') \, .
\end{equation}
This expression involves the (angular diameter) distances $D_{LS}$ (from the lens to the source), $D_L$ (from the observer to the lens) and $D_S$ (from the observer to the source).\footnote{
    We compute these with \texttt{astropy}~\citep{astropy:2013,astropy:2018} using the flat cosmology from Planck \citep{Planck2018}.
} Since lensing merely alters the trajectories of photons rather than creating or destroying them, the surface brightness $B(\bm{\xi})$ in the image plane is equal to the surface brightness at the point to which it traces back in the source plane:
\begin{equation}
    B(\bm{\xi}) = B(\bm{x}(\bm{\xi})) \, .
\end{equation}
Our lens model thus requires specifying the form of the deflection fields of lens components and the surface brightness of the source.

\subsection{Source}

The brightness profile of our mock sources is parametrized by the widely-used Sérsic profile:
\begin{equation}
    f(\bm{x}) = I_e \exp\left\{ -k_n \left[ \left( \frac{R(\bm{x})}{r_e} \right)^{1 / n} - 1 \right] \right\} \, ,
\end{equation}
where $r_e$ is the half-light radius and $k_n$ is a normalization constant related to the index $n$. For $n > 0.36$ \citep{Ciotti:1999zs}
\begin{equation}
    k_n \approx 2 n - \frac{1}{3} + \frac{4}{405 n} + \frac{46}{25515 n^2} + \frac{131}{1148175 n^3} - \frac{2194697}{30690717750 n^4} \, .
\end{equation}
For typical galaxies $1/2 < n < 10$. The radial parameter $R(\bm{x})$ is the length of the elliptical coordinate vector
\begin{equation}
    \begin{pmatrix} R_x \\ R_y \end{pmatrix} = \begin{pmatrix} q^{1/2} & 0 \\ 0 & q^{-1/2} \end{pmatrix} \begin{pmatrix} \cos\varphi & \sin\varphi \\ -\sin\varphi & \cos\varphi \end{pmatrix} \begin{pmatrix} x - x_0 \\ y - y_0 \end{pmatrix} \, ,
\end{equation}
which depends on the source's position angle $\varphi$, axis ratio $q$ and position $(x_0, y_0)$. We fix the source's redshift to $z_\mathrm{src} = 2$.

Our source model therefore has seven parameters, $\psrc \equiv (x_\mathrm{src}, y_\mathrm{src}, \varphi_\mathrm{src}, q_\mathrm{src}, n, r_e, I_e)$.

\subsection{Main lens}
\label{sec:sple}

We adopt the \gls*{sple} model for the main lens galaxy, which is capable of modeling the gravitational potentials of strong lenses to near the percent level~\citep{Suyu:2008zp}. The \gls*{sple} deflection field can be expressed in closed-form as a complex field $\alpha = \alpha_x + i \alpha_y$ \citep{Tessore:2015baa,2020MNRAS.496.3424O}:
\begin{equation}
\begin{split}
    \bm{\alpha}^\mathrm{SPLE}(\bm{\xi}) = \theta_E & \frac{2 q_\mathrm{lens}^{1/2}}{1 + q_\mathrm{lens}} \left( \frac{\theta_E}{R} \right)^{\gamma - 2} e^{i \phi} \\
    & \cdot \hypgeom\left( 1, \frac{\gamma - 1}{2}, \frac{5 - \gamma}{2}, -\frac{1-q_\mathrm{lens}}{1+q_\mathrm{lens}} e^{2 i \phi} \right) \, .
\end{split}
\end{equation}
Here $(R, \phi)$ are elliptical coordinates, related to the Cartesian coordinates $\bm{\xi}$ through a transformation parametrized by the lens' orientation $\varphi_\mathrm{lens}$, axis ratio $q_\mathrm{lens}$ and position $(x_\mathrm{lens}, y_\mathrm{lens})$:
\begin{align}
    \begin{pmatrix} R_x \\ R_y \end{pmatrix} &= \begin{pmatrix} q_\mathrm{lens}^{1/2} & 0 \\ 0 & q_\mathrm{lens}^{-1/2} \end{pmatrix} \begin{pmatrix} \cos\varphi_\mathrm{lens} & \sin\varphi_\mathrm{lens} \\ -\sin\varphi_\mathrm{lens} & \cos\varphi_\mathrm{lens} \end{pmatrix} \begin{pmatrix} \xi_x - x_\mathrm{lens} \\ \xi_y - y_\mathrm{lens} \end{pmatrix} \, ,\\
    \tan \phi &= \frac{R_y}{R_x}.
\end{align}
Since the hypergeometric function $\hypgeom$ is not implemented in \texttt{PyTorch}, we instead pretabulate its value as a function of $\phi$, $q_\mathrm{lens}$ and $\gamma$ and interpolate, as described in \citet{Chianese:2019ifk}.

The slope $\gamma$ has a complicated degeneracy with the size of the source \citep{Schneider:2013sxa,Schneider:2013wga}. Roughly, larger $\gamma$ values cause the spatial scale of the source to increase \citep[sec. 3.3]{2015MNRAS.452.2940N}. For simplicity we fix $\gamma = 2.1$. We also assume the lens galaxy's light has been perfectly subtracted, and fix its redshift to $z_\mathrm{lens} = 0.5$.

To account for the weak lensing due to large-scale structure located along the line of sight to the source, we also include an external shear component, which is constant across the image plane:
\begin{equation}
    \bm{\alpha}^\mathrm{shear}(\bm{\xi}) = \begin{pmatrix} \gamma_1 & \gamma_2 \\ \gamma_2 & -\gamma_1 \end{pmatrix} \bm{\xi} \, .
\end{equation}

Our main lens model thus has seven parameters: the \gls*{sple} parameters $(x_\mathrm{lens}, y_\mathrm{lens}, \varphi_\mathrm{lens}, q_\mathrm{lens}, \theta_E)$ and the external shear parameters $(\gamma_1, \gamma_2)$, which we denote collectively with $\plens$.

\subsection{Perturbers}
\label{sec:perturbers}

\subsubsection{Density profiles}

We model the deflection field of subhalos using a truncated \gls*{nfw} profile \citep{Baltz_2009} to account for tidal stripping by the main lens:
\begin{equation}
    \rho_\mathrm{NFW}(r) = \frac{\rho_s}{r/r_s \, (1 + r/r_s)^2} \, \frac{1}{1 + r^2/r_t^2} \, .
\end{equation}
Here $r$ is the distance from the center of the subhalo, $\rho_s$ is the density normalization, $r_s$ is the scale radius and $r_t$ is the truncation radius. The deflection field for this density profile is given in \citet[appendix A]{Baltz_2009}, and differs from that of an \gls*{nfw} profile for $r \gtrsim r_t$. While the value of $\tau \equiv r_t / r_s$ depends on the full history of the subhalo, it typically falls between \num{4} and \num{10}~\citep{Gilman:2019nap}; we fix $\tau = 6$ for simplicity. For simplicity, we model \gls*{los} halos using exactly the same profile even though they typically have not undergone tidal stripping.

To generate perturber populations for our third analysis task, we must choose values for their density normalizations and scale radii. Since simulation studies typically measure the halo mass $m_\mathrm{sub}$\footnote{
    This is defined as the mass of the halo enclosed in a sphere where the untruncated halo's average density is 200 times the critical density.
} and the concentration $c$, it is more convenient to sample populations from distributions over these parameters. These variables can then be mapped to the parametrization above via
\begin{align}
    r_s &= \frac{1}{c} \left[ \frac{3 m_\mathrm{sub}}{4 \pi \, 200 \rho_\mathrm{cr}(z_\mathrm{lens})} \right]^{1/3} \, ,\\
    \rho_s &= \rho_\mathrm{cr}(z_\mathrm{lens}) \, \frac{1}{3} \frac{c^3}{\log(1+c) - c / (1 + c)} \, . 
\end{align}
For simplicity we fix $c = 15$, which is roughly the average value for perturbers in the mass range \SIrange{e7}{e10}{\solarmass} \citep[fig. 7]{Richings:2020auv}. We anticipate that accounting for scatter in the mass-concentration relation might actually improve our ability to measure subhalos' parameters as higher concentrations lead to substantially stronger lensing signals \citep{Amorisco:2021iim}.

The parameters of an individual subhalo which are not fixed are thus $\psub \equiv (x_\mathrm{sub}, y_\mathrm{sub}, m_\mathrm{sub})$, where the second and third components are the projected position of the subhalo. In the case of \gls*{los} halos, the parameter set also includes the redshift $z_\mathrm{los}$. In the next two subsections, we describe how we sample these parameters.

\subsubsection{Generating subhalos}

We sample subhalo masses using a mass function of the form from \citet{2010MNRAS.404..502G}:
\begin{equation}
    \dv{n}{\log m_{200}} = m_{200} (1 + z_\mathrm{lens})^{1/2} A_M m_{200}^{-\alpha} \exp\left[ -\beta \left( \frac{m_{200}}{M_{200}} \right)^3 \right] \, ,
\end{equation}
where $M_{200}$ is the mass of the main lens. The free parameters in this function were fit to hydrodynamical cosmological simulations that included baryons in \citet{Despali:2016meh}. In particular, we use the fits to EAGLE, which give $\alpha = 0.85$ (given in the text) and $(A_M, \beta) = (\num{2.4e-4}\, \msun^{\alpha - 1}, 300)$ (extracted from their figures). Integrating the mass function over a given mass interval gives the expected number of subhalos in that interval distributed throughout the whole main lens.

\citet{Despali:2016meh} found the distribution of radial coordinates in hydrodynamical simulations is well-fit by an Einasto profile, but can be approximated as uniform over the lens plane. For a given lensing system we thus precompute $\bar{n}_\mathrm{sub}$, the number of subhalos expected to fall within the lens plane. Thereafter we generate the subhalo population by sampling the number of subhalos from $\operatorname{Poisson}(\bar{n}_\mathrm{sub})$, drawing their masses from the subhalo mass function and sampling their projected positions uniformly over the lens plane. Since the vast majority of subhalos fall outside the lens plane, we expect their lensing effect to be mostly degenerate with external shear, and thus do not simulate them. With the lens redshift we have chosen, over a $\SI{5}{\arcsecond} \times \SI{5}{\arcsecond}$ image and integrating over the mass range \SIrange{e7}{e8}{\solarmass}, we find $\bar{n}_\mathrm{sub} = 3.1$.

\subsubsection{Generating line-of-sight halos}

As described in \citet{Montel:2022fhv,CaganSengul:2020nat}, we first compute the average number of \gls*{los} halos in the double-pyramid geometry connecting the observer, lens-plane and source. For each simulation we sample the number of \gls*{los} halos from $\operatorname{Poisson}(\bar{n}_\mathrm{los})$. We then sample their redshifts and projected positions uniformly over the double-pyramid region and draw their masses from the mass function in \citet{Tinker:2008ff}, with $\Delta$ set to \num{200}. For the lens and source redshifts we have chosen, $\bar{n}_\mathrm{los} = 265.6$.

To avoid expensive iterative ray-tracing through the lens planes of each \gls*{los} halo, we project them as effective subhalos into the lens plane, using the relations derived in \citet{CaganSengul:2020nat} to rescale their scale radii and masses. Performing the full multiplane ray-tracing would likely lead to slightly tighter posteriors for the subhalo parameters we infer in this work, as the multiplane deflection field generally has a small curl component that cannot be mimicked by single-plane lensing. As with subhalos, we ignore any \gls*{los} halos lying outside the double pyramid volume.

\subsection{Instrumental effects}

We generate mock data with comparable quality to Hubble Space Telescope observations. All images are $\SI{5}{\arcsecond} \times \SI{5}{\arcsecond}$ with $\SI{0.05}{\arcsecond}$ resolution ($100 \times 100$ pixels). We do not include a \gls*{psf}. To account for the fact that each pixel in the image corresponds to a finite collecting region in the sky, we generate our images at a resolution eight times higher than the target resolution and downsample. In experiments we found that neglecting this effect can have a significant impact on inference results. Lastly, we add Gaussian pixel noise to our observations such that the brightest pixels are approximately 30 times the noise level.

\section{Inference with truncated marginal neural ratio estimation}
\label{sec:inference}

In the inference tasks we confront in the rest of this work, our goal is to infer 2D marginals for the position and 1D marginals for the mass of a subhalo. Each posterior is to be marginalized over the other perturber parameters and potentially another set of parameters $\bm{\eta}$ for the main lens, source and perturber population. In this section we review how \gls*{tmnre} solves such inference problems.

To begin with, \gls*{nre} \citep{aalr} is a technique for inferring the posterior $p(\vartheta | \data)$ for a model with the joint distribution $p(\data, \vartheta)$, where $\data$ is an observation (e.g. a lensing image) and $\vartheta$ is a parameter of interest (e.g. the mass of a subhalo). The idea is to train a classifier to distinguish between data and parameters drawn from two classes labeled by the binary variable $C$:
\begin{align}
    p(\data, \vartheta | C = 0) &= p(\data) p(\vartheta) \\
    p(\data, \vartheta | C = 1) &= p(\data, \vartheta) \, .
\end{align}
These two distributions correspond respectively to simulating data from the simulator and drawing an unrelated set of parameters from the prior versus sampling parameters and data from the simulator. Sampling $C=0$ and $C=1$ with equal probability, the decision function for the (Bayes-)optimal classifier can be computed using Bayes' theorem:
\begin{equation}
    p(C = 1 | \data, \vartheta) = \frac{p(\data, \vartheta)}{p(\data, \vartheta) + p(\data) p(\vartheta)} \equiv \sigma[ \log r(\data, \vartheta)] \, ,
\end{equation}
where we introduced the sigmoid function $\sigma(y) \equiv 1 / (1 + e^{-y})$ and the likelihood-to-evidence ratio:
\begin{equation}
    r(\data,\vartheta) \equiv \frac{p(\data | \vartheta)}{p(\data)} = \frac{p(\data, \vartheta)}{p(\data) p(\vartheta)} =  \frac{p(\vartheta | \data)}{p(\vartheta)} \, .
\end{equation}
Therefore, by training a neural network $\hat{r}_\phi(\data, \vartheta)$ to estimate $r(\data, \vartheta)$ via this supervised classification task,\footnote{
    For better numerical stability, we actually train the network to learn $\log r(\data, \vartheta)$.
} we obtain an estimate of the posterior through $\hat{p}_\phi(\vartheta | \data) = \hat{r}_\phi(\data, \vartheta) p(\vartheta)$. This \emph{ratio estimator} can be trained by minimizing the binary cross-entropy loss
\begin{equation}
\label{eq:bce-loss}
\begin{split}
    \ell[\hat{r}_\phi] &= -\int \dd{\data} \dd{\vartheta} \left\{ p(\data, \vartheta) \log \sigma[ \log \hat{r}_\phi(\data, \vartheta) ] \right. \\
    &\hspace{2cm} \left. + p(\data) p(\vartheta) \log\left[ 1 - \sigma[ \log \hat{r}_\phi(\data, \vartheta) ] \right] \right\}
\end{split}
\end{equation}
with respect to the ratio estimator's parameters $\phi$ using stochastic gradient descent techniques. Critically, training only requires the ability to generate \emph{samples} from the simulator. This makes it straightforward to apply marginal ratio estimation in scenarios where the explicit form of the likelihood cannot be written in closed form. In practice, posterior samples can be generated by resampling prior samples (with replacement) weighted by the ratio, enabling posterior sampling even when the prior cannot be expressed in closed form.

The extension to estimating marginal posteriors is straightforward: parameters to be marginalized over must be sampled, but \emph{not} presented to the ratio estimator. In more detail, consider a model with the joint distribution $p(\data, \bm{\eta}, \vartheta) = p(\data | \bm{\eta}, \vartheta) p(\bm{\eta}, \vartheta)$, where $\bm{\eta}$ is a set of parameters to be marginalized over (e.g. the source and main lens parameters). If $\bm{\eta}$ is not passed to the ratio estimator, the loss function becomes
\begin{align}
    \ell[\hat{r}_\phi] &= -\int \dd{\data} \dd{\vartheta} \dd{\bm{\eta}} \left\{ p(\data, \bm{\eta}, \vartheta) \log \sigma[ \log \hat{r}_\phi(\data, \vartheta) ] \right. \notag \\
    &\hspace{1.8cm} \left. + p(\data) p(\bm{\eta}, \vartheta) \log\left[ 1 - \sigma[ \log \hat{r}_\phi(\data, \vartheta) ] \right] \right\} \label{eq:bce-with-nuisance} \\
    &= -\int \dd{\data} \dd{\vartheta} \left\{ p(\data, \vartheta) \log \sigma[ \log \hat{r}_\phi(\data, \vartheta) ] \right. \notag \\
    &\hspace{1.8cm} \left. + p(\data) p(\vartheta) \log\left[ 1 - \sigma[ \log \hat{r}_\phi(\data, \vartheta) ] \right] \right\}
\end{align}
where we integrated over $\bm{\eta}$ to obtain the second equality, proving our statement.

The ratio estimators discussed so far are fully \emph{amortized}: that is, they attempt to learn $r(\data, \vartheta)$ over the whole range of the prior $p(\bm{\eta}, \vartheta)$. In principle, it is useful to be able to analyze any possible observation with the same network. In practice, when the posterior $p(\bm{\eta}, \vartheta | \data_0)$ for a particular observation $\data_0$ is much narrower than the prior, training an accurate ratio estimator requires a massive amount of training data. We instead focus on the problem of \emph{targeted inference} of the posterior for $\data_0$, which substantially reduces training data requirements and reduces the complexity of the function the ratio estimator must learn to model. Such an approach is also well-suited to individually targeting the small sample of lenses relevant to \gls*{dm} substructure measurement that exist at present ($\mathcal{O}(100)$).

Training targeted ratio estimators is achieved by replacing the prior with a \emph{truncated prior} $p_\Gamma(\bm{\eta}, \vartheta)$, where the parameters are restricted to a region $\Gamma$ where they are likely to have generated $\data_0$. Since parameters from the complement of $\Gamma$ are unlikely to have generated $\data_0$, training a ratio estimator with data generated from the truncated prior as opposed to the full prior has little impact on the posterior learned by our ratio estimators.

Since the highest probability density region of the true posterior $\Gamma$ is unknown, we compute an estimate $\hat{\Gamma}$ over a sequence of inference rounds. At the beginning of each round, we sample from $p_{\hat{\Gamma}}(\bm{\eta}, \vartheta)$ (or the true prior in the first round) and train a ratio estimator. We re-estimate $\hat{\Gamma}$ by keeping only the parts of the previous truncated prior for which $\hat{r}_\phi(\data, \vartheta)$ exceeds a certain threshold, as described in \citet{tmnre}. This determines the truncated prior for the next round. Our final ratio estimator is obtained when $\hat{\Gamma}$ stops changing substantially between rounds. This whole procedure is called \gls*{tmnre}.

\gls*{tmnre} is related to other \emph{sequential} \gls*{sbi} methods, such as sequential neural posterior estimation \citep{2016arXiv160506376P,DBLP:journals/corr/abs-1905-07488,2017arXiv171101861L} and sequential neural likelihood estimation \citep{2018arXiv180507226P}. These two methods use the \emph{posteriors} learned in each round to generate simulations for the next round rather than the truncated prior. This approach is inefficient for learning multiple marginal posteriors simultaneously, since sampling from the marginal for a particular parameter may hinder learning the marginals for other parameters.

The fact that \gls*{tmnre} learns a function that can be rapidly evaluated makes it possible to perform statistical consistency checks. In this work we perform expected coverage checks \citep{Cole:2021gwr,averting} to test the calibration of our ratio estimators for observations generated using parameters from the truncated prior. This test measures whether credible regions of different widths have achieved their nominal coverage (i.e., whether the true parameters fall within the 68\% credible interval of the estimated posterior for 68\% of observations). Agreement between the nominal and empirically-measured expected coverage is a necessary (but not sufficient) condition for the ratio estimator to be a correct estimate of the posterior. While typically expected coverage tests are a statement about the ratio estimator's properties averaged over the truncated prior, at increased computational cost the coverage can be checked in a frequentist manner as a function of the true parameters.

\section{Results}
\label{sec:results}

We now apply \gls*{tmnre} to three different substructure lensing problems of increasing complexity. For all tasks we use the same general ratio estimator architecture. It consists of an initial compression network that maps the $100 \times 100$ pixel images into a feature vector. This feature vector is concatenated to $\psub$ (and separately to $\psrc$ and $\plens$ for tasks where they are also inferred). The vector is then passed to a \gls*{mlp} which outputs an estimate of the 2D and 1D marginal likelihood-to-evidence ratios for $(x_\mathrm{sub}, y_\mathrm{sub})$ and $\log_{10} m_\mathrm{sub} / \msun$ respectively (with separate \gls*{mlp}s used to estimate the 1D ratios for $\psrc$ and $\plens$).

For each ratio estimator we begin the first training round with \num{10000} training examples. We then truncate each parameter's prior. If none of the truncated priors shrank by at least 20\%, we increase the number of training examples by a factor of \num{1.5} for the next inference round. A fresh network is then trained using simulations drawn from the truncated prior. Convergence of the ratio estimator is declared after five such consecutive increases in the training set size. For tasks in which we must infer the macromodel parameters we first train the macromodel ratio estimator using this procedure and use the resulting truncated priors to generate training data for the subhalo ratio estimators using the same training procedure. We use the implementation of \gls*{tmnre} in \texttt{swyft}\footnote{
    \url{https://github.com/undark-lab/swyft/}
} \citep{swyftjoss}, which is built on \texttt{PyTorch} and \texttt{pytorch-lightning}\footnote{
    \url{https://www.pytorchlightning.ai/}
}.

The training data for our ratio estimators differs in important ways from typical datasets studied by machine learning researchers, making the choice of a good compression network an interesting challenge. Consider, for example, the machine learning problem of classifying the content of natural images. Natural images are distinguished by a hierarchy of visual features at different scales (for example, small-scale features such as textures and edges which comprise large-scale features like the head of an animal or part of an object). A good image classifier should be translation-invariant, producing the same output regardless of the position of an image's contents. Since the deep \gls*{cnn} architecture has an inductive bias towards learning a hierarchy of features and are translation invariant, \gls*{cnn}s are widely used in computer vision.

The training data for our ratio estimators does not share these features. Different perturber configurations produce images with slightly different relationships between the multiple images of the source galaxy. The variations between images lie near the Einstein ring, and do not show the same rich hierarchical structure of natural images. This means that inductive biases of \gls*{cnn}s are not necessarily beneficial in the context of substructure lensing.

In our experiments, we used \gls*{cnn}s in the ratio estimators for the macromodel parameters, finding their performance to be adequate. However, we found they produced much too wide 2D marginals for the position of a subhalo. Instead, we found the MLP Mixer \citep{mlpmixer} to work well.\footnote{
    The MLP Mixer implementation we use can be found at \url{https://github.com/lucidrains/mlp-mixer-pytorch}.
} Roughly, the MLP Mixer splits the image into patches, stacks the patches and passes each pixel in the stack through an \gls*{mlp}, acting as a dilated convolution. (Another MLP is then applied along the channel dimension of the mixed patches, and the process is iterated.) The MLP Mixer thus directly examines the relationships between pixels in disparate parts of the image, which is exactly how the properties of subhalos are imprinted. We expect that other architectures that split the image into patches such as \gls*{vit} \citep{vit} could work well for the compression network, though \gls*{vit} is known to require large amounts of training data.

The architectures of our macromodel and subhalo compression networks are given in \cref{app:architectures}. While we did not perform a full hyperparameter exploration, we found the batchnorm layers to be crucial for stable training of the \gls*{cnn} used for the macromodel ratio estimator. Since our images are roughly one-quarter the area of the images studied in the paper introducing MLP Mixer, we use a substantially smaller model than they suggest. Using dropout in the MLP Mixer and classifier \gls*{mlp}s improved performance. Varying the number of hidden layers and their size in the classifiers had little impact.

We used the Adam optimizer with an initial learning rate of \num{6e-3} for the macromodel ratio estimator and \num{4e-4} for the subhalo ratio estimator (found through a learning rate test) and a batch size of \num{64}. The learning rate was reduced by a factor of \num{0.1} whenever the validation loss plateaued for \num{3} epochs. Training was run for no longer than \num{30} epochs.

\begin{table}
    \caption{True subhalo and macromodel parameter values and priors used in the first \gls*{tmnre} inference round in our three inference tasks. The last column references the first section in which the indicated parameter is inferred rather than being fixed to its true value. The slope of the main lens is fixed to \num{2.1}, as explained in \ref{sec:sple}. The main lens and source redshifts are set to $z_\mathrm{lens} = 0.5$ and $z_\mathrm{src} = 2$ respectively. For the analysis in \cref{sec:lshs} involving a population of light perturbers, we sample the number of \gls*{los} and subhalos from Poisson distributions with means $\bar{n}_\mathrm{los} = 265.6$ and $\bar{n}_\mathrm{sub} = 3.1$ respectively, and restrict their masses to the range \SIrange{e7}{e8}{\solarmass}. The halo mass functions and redshift distributions are described in detail in \cref{sec:perturbers}. For all perturbers we fix the concentration to $c = 15$ and truncation scale $\tau = r_t / r_s = 6$.}
    \centering
    \begin{tabular}{c c c c c}
        \toprule
        & Parameter & True value & Initial prior & First inferred in \\
        \midrule
        \parbox[t]{1mm}{\multirow{3}{*}{\rotatebox[origin=c]{90}{Subhalo}}}
        & $x_\mathrm{sub}\, ['']$ & \num{-1.1} & $\uniform(-2.5, 2.5)$ & \cref{sec:sp} \\
        & $y_\mathrm{sub}\, ['']$ & \num{-1.1} & $\uniform(-2.5, 2.5)$ & \cref{sec:sp} \\
        & $\log_{10} m_\mathrm{sub}/\msun$ & \num{9.5} & $\uniform(8, 10.5)$ & \cref{sec:lss}\\
        \midrule
        \parbox[t]{1mm}{\multirow{6}{*}{\rotatebox[origin=c]{90}{SPLE}}}
        & $x_\mathrm{lens}\, ['']$ & -0.05 & $\uniform(-0.2, 0.2)$ & \multirow{6}{*}{\cref{sec:lss}} \\
        & $y_\mathrm{lens}\, ['']$ & 0.1 & $\uniform(-0.2, 0.2)$ \\
        & $\varphi_\mathrm{lens} \, [^\circ]$ & 1 & $\uniform(0.5, 1.5)$ \\
        & $q_\mathrm{lens}$ & 0.75 & $\uniform(0.5, 1)$ \\
        & $\gamma$ & 2.1 & --- \\
        & $r_\mathrm{ein}\, ['']$ & 1.5 & $\uniform(1, 2)$ \\
        \midrule
        \parbox[t]{1mm}{\multirow{2}{*}{\rotatebox[origin=c]{90}{Shear}}}
        & $\gamma_1$ & 0.005 & $\uniform(-0.5, 0.5)$ & \multirow{2}{*}{\cref{sec:lss}} \\
        & $\gamma_2$ & -0.010 & $\uniform(-0.5, 0.5)$ \\
        \midrule
        \parbox[t]{1mm}{\multirow{7}{*}{\rotatebox[origin=c]{90}{Source}}}
        & $x_\mathrm{src}\, ['']$ & 0 & $\uniform(-0.2, 0.2)$ & \multirow{7}{*}{\cref{sec:lss}} \\
        & $y_\mathrm{src}\, ['']$ & 0 & $\uniform(-0.2, 0.2)$ \\
        & $\varphi_\mathrm{src} \, [^\circ]$ & 0.75 & $\uniform(0.5, 1.25)$ \\
        & $q_\mathrm{src}$ & 0.5 & $\uniform(0.2, 0.8)$ \\
        & $n$ & 2.3 & $\uniform(1.5, 3)$ \\
        & $r_e\, ['']$ & 2.0 & $\uniform(0.5, 3)$ \\
        & $I_e$ & 0.6 & $\uniform(0.1, 2)$ \\
        \bottomrule
    \end{tabular}
    \label{tab:params-and-priors}
\end{table}

\subsection{Subhalo position inference with fixed mass, source and lens}
\label{sec:sp}

We first consider the case where the only free parameters in the lens are the position of a single \SI{e9}{\solarmass} subhalo, $\psub = (x_\mathrm{sub}, y_\mathrm{sub})$. The prior is taken to be uniform over the image plane (i.e., $\uniform(-2.5, 2.5)$ for both coordinates). The posterior for $\psub$ can then be computed analytically. Adopting a uniform prior over $\psub$ covering the image plane and using the fact the posterior is much narrower, we have
\begin{equation}
    \log p(\psub | \data) 
    \sim -\frac{1}{2} \sum_{i,j} \left( \frac{x_{ij} - f_{ij}(\psub)}{\sigma_n} \right)^2 \, ,
\end{equation}
where the sum runs over pixels and we dropped terms independent of $\psub$.

\begin{figure*}
    \centering
    \includegraphics[height=7cm]{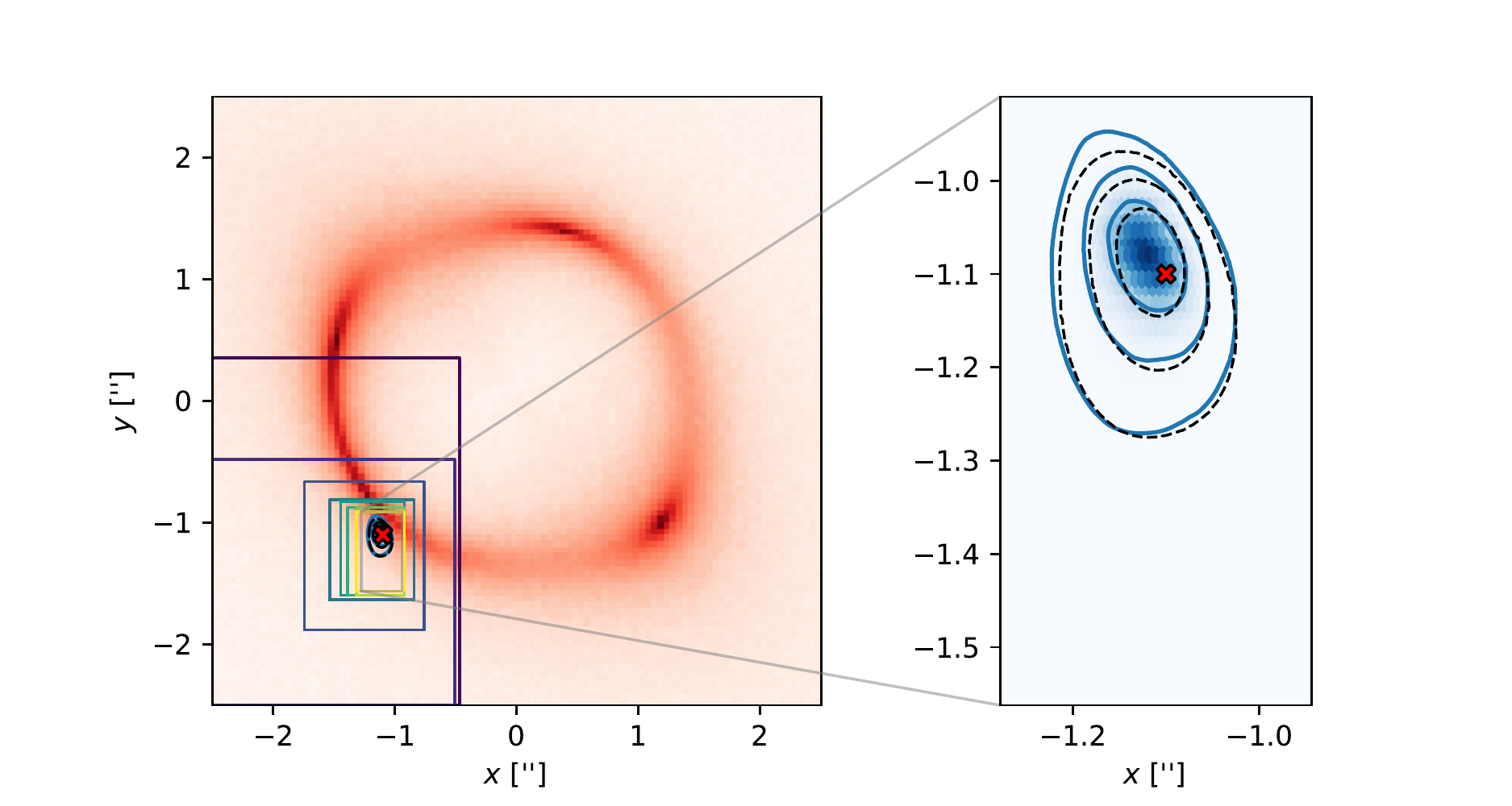}
    \caption{Validation of \gls*{tmnre} through inference of the position of a subhalo, with macromodel parameters fixed to their true values and the subhalo's mass fixed to \SI{e9}{\solarmass}. The observation is shown in the left panel. The blue and dashed black contours correspond to the posterior inferred with \gls*{tmnre} and computed analytically respectively, indicating the 68\%, 95\% and 99.7\% credible regions. The red {\color{red} $\times$} shows the subhalo's true position. The blue through yellow boxes in the left panel show the ranges of the truncated prior based on the 1D marginals for the subhalo's coordinates. The zoom-in on the right encompasses the range of the final truncated prior. The distorted blue hex-bin histogram shows the magnitude of the inferred posterior.}
    \label{fig:sp-post}
\end{figure*}

\begin{figure}
    \centering
    \includegraphics[height=7cm]{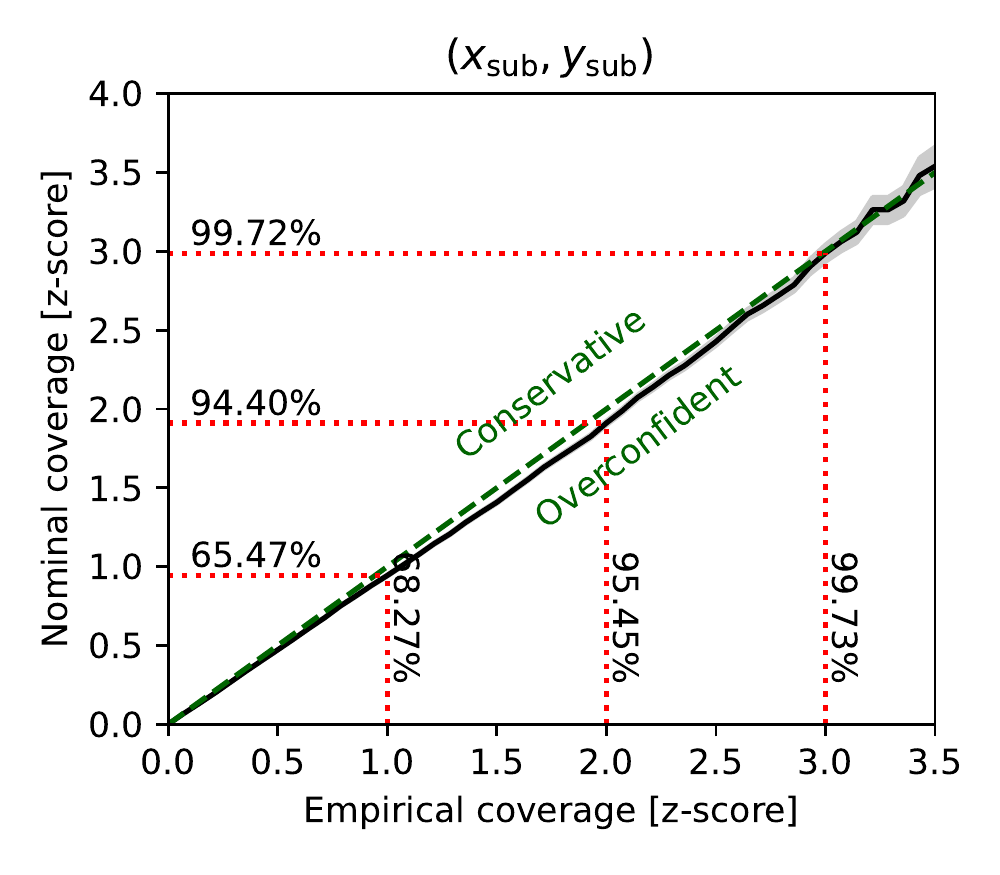}
    \caption{Coverage plot for inference task where only the subhalo's position is free (see \cref{fig:sp-post}), showing our ratio estimator produces posteriors of the correct size on average. In detail, the black curve shows the empirical versus nominal coverage, estimated by computing posteriors for \num{10000} observations drawn from the final truncated prior. The statistical uncertainty of this estimate is plotted in grey; its derivation is explained in detail in \citet{Cole:2021gwr}. For a perfectly-calibrated ratio estimator, the black curve would lie along the diagonal green dashed line. The red dashed lines indicate the empirical and nominal coverage of the $1 - 3 \sigma$ credible regions.}
    \label{fig:sp-coverage}
\end{figure}

\Cref{fig:sp-post} shows the truncation regions for each round and compares the analytically-computed posterior with the posterior inferred using \gls*{tmnre}. While the truncation regions and posterior estimates in early rounds are extremely broad compared to the analytically-computed posterior, \gls*{tmnre} successfully identifies the region of the image containing the subhalo. After 10 inference rounds the truncation region stabilizes and the inferred posteriors agree well with the true ones for each coordinate. To complement this visual check we also check the coverage for samples from the final round of \gls*{tmnre} in \cref{fig:sp-coverage}. We find the empirical and nominal coverage to be in good agreement, with our ratio estimator very slightly underestimating the width of the posterior beyond the 95\% confidence level. 

Having validated \gls*{tmnre} in this simple scenario, we now turn to more complex inference tasks where the posteriors of interest cannot be derived analytically.


\subsection{Subhalo mass and position inference}
\label{sec:lss}

Next we aim to infer the position and mass of a single subhalo, $\psub = (m_\mathrm{sub}, x_\mathrm{sub}, y_\mathrm{sub})$, in a system where the source and main lens parameters are also unknown. The priors for the \num{17} parameters of the model are given in \cref{tab:params-and-priors}. Due to the relatively low dimensionality, inference on this model is within the reach of likelihood-based tools such as \gls*{mcmc} or nested sampling. In addition, it can be implemented in a differentiable manner, making the application of methods such as \gls*{hmc} possible \citep{Gu:2022xhk,Chianese:2019ifk}. Running such expensive scans is beyond the scope of this paper.

\begin{figure*}
    \centering
    \includegraphics[height=6cm]{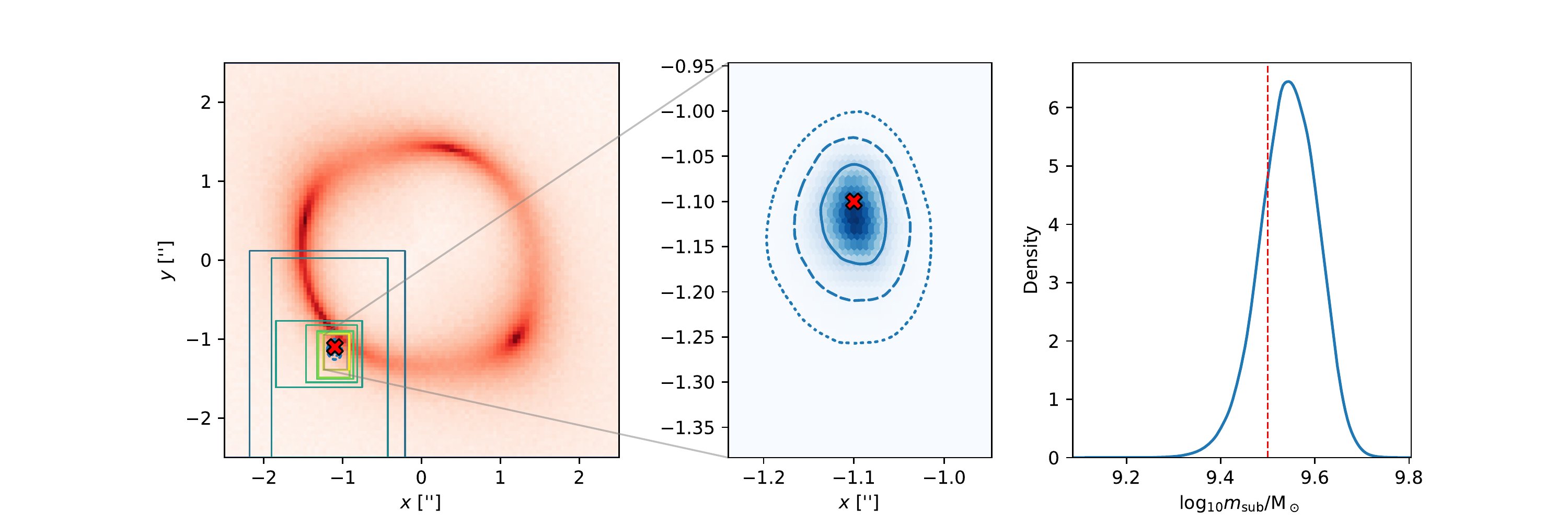}
    \caption{Marginal posteriors inferred with \gls*{tmnre} for a subhalo's 2D position (left and center) and mass (right) in a lens with unknown macromodel parameters. See the caption of \cref{fig:sp-post} for further details, though note we have instead used solid, dashed and dotted lines respectively to mark the 68\%, 95\% and 99.7\% credible regions of the position posterior. The range of the $x$-axis in the right panel shows the final-round truncated prior for the subhalo's $\log_{10}$-mass.}
    \label{fig:lss-sub-post}
\end{figure*}

\begin{figure*}
    \centering
    \includegraphics[width=0.84\linewidth]{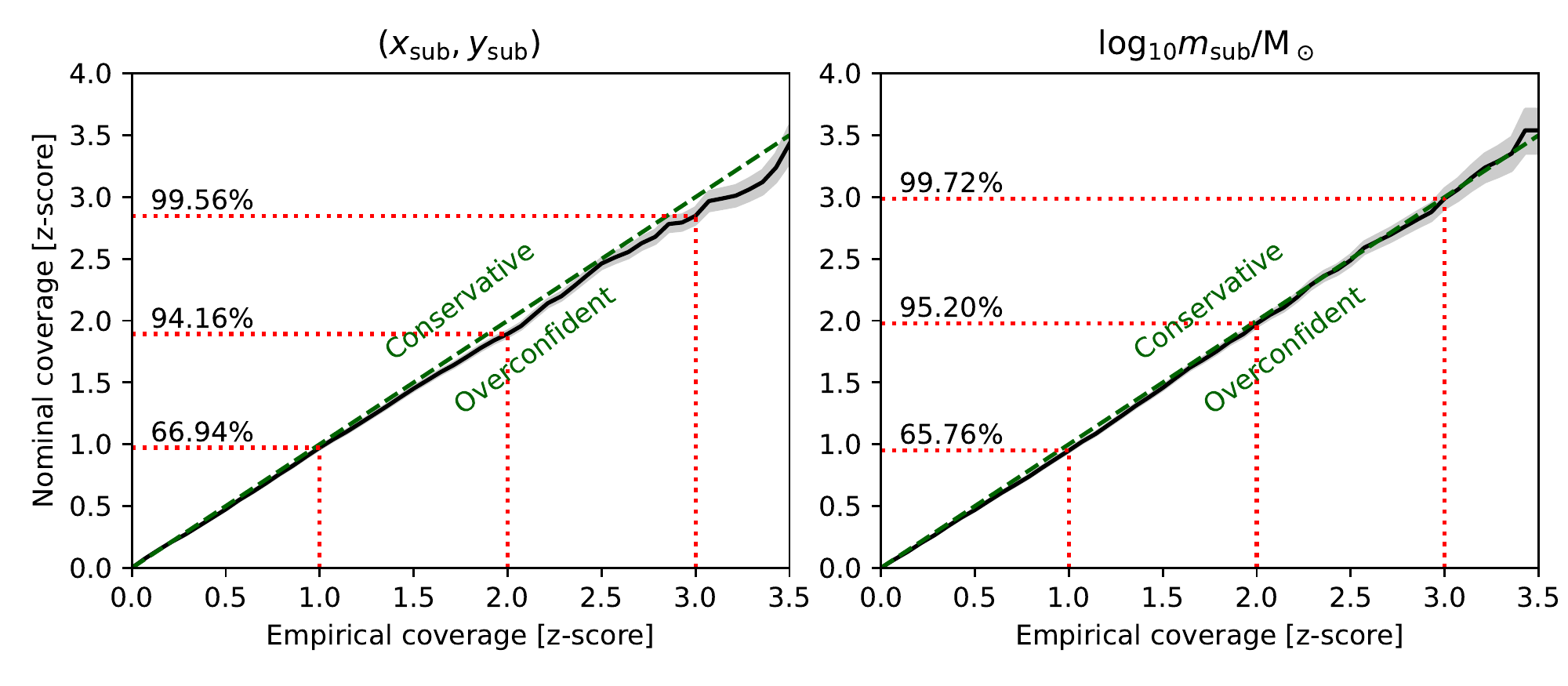}
    \caption{Coverage plots for subhalo position and mass ratio estimators learned from the observation in \cref{fig:lss-sub-post}. These again indicate the estimators' credible regions are on average the correct size for observations drawn from the final-round truncated prior. See \cref{fig:sp-coverage} for an explanation of the format.}
    \label{fig:lss-sub-coverage}
\end{figure*}

\begin{figure*}
    \centering
    \includegraphics[width=0.84\linewidth]{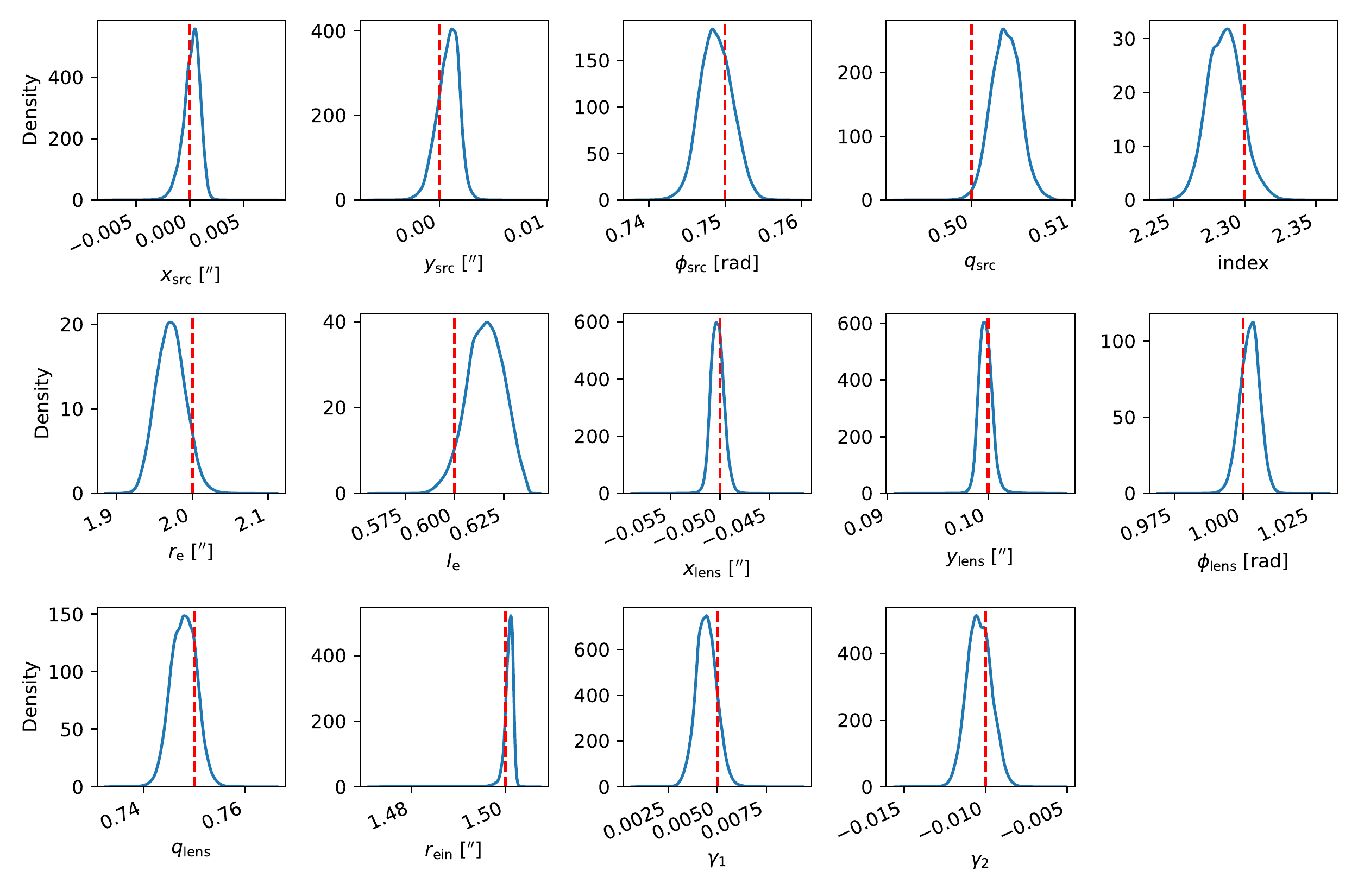}
    \caption{The 1D marginal posteriors for all macromodel parameters of the lensing system shown in \cref{fig:lss-sub-post}. The posteriors were computed using a \gls*{cnn}-based ratio estimator. The first seven panels correspond to the source parameters, the next five are for the main lens and the last two are for the external shear. All posteriors encompass the true parameter values (vertical red dashed lines) within the $\sim 2\sigma$ interval.}
    \label{fig:lss-macro-post}
\end{figure*}

\begin{figure*}
    \centering
    \includegraphics[width=0.84\linewidth]{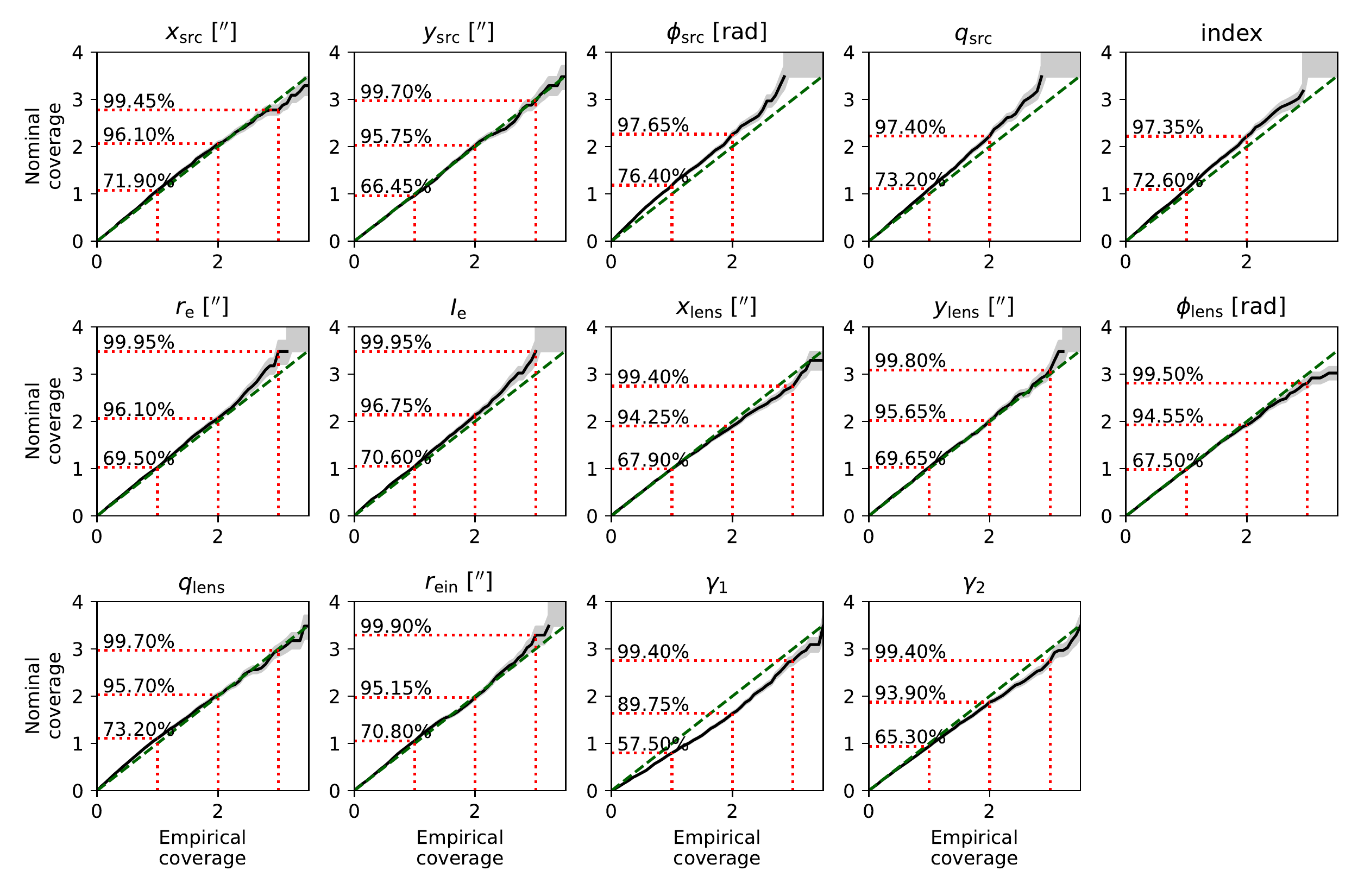}
    \caption{Coverage plots for the 1D marginal macromodel parameter posteriors of the lensing system from \cref{fig:lss-sub-post}, using the same format as in \cref{fig:sp-coverage}. The posteriors generally have coverage, with a few being slightly conservative ($\phi_\mathrm{src}$, $q_\mathrm{src}$ and the source index) and the shear posteriors being slightly overconfident.}
    \label{fig:lss-macro-coverage}
\end{figure*}

The final posteriors for the subhalo parameters are shown in \cref{fig:lss-sub-post}. The true values of all parameters fall within the $\sim 68\%$ credible intervals of the inferred posteriors. We find the effect of the uncertain macromodel is not too strong (at least for this noise realization), with the size of the subhalo position posterior being comparable to what we found in the previous inference task.
\Cref{fig:lss-sub-coverage} demonstrates that our ratio estimator has good coverage with respect to the constrained prior. In \cref{fig:lss-macro-post,fig:lss-macro-coverage} we display the marginal posteriors and coverage plots for all 14 source and main lens parameters, which demonstrate they are well-calibrated.


\subsection{Subhalo mass and position inference with a population of perturbers}
\label{sec:lshs}

For our final inference task we extend the previous one by aiming to infer the position and mass of a relatively heavy target subhalo while marginalizing over a population of lighter perturbers of unknown size. The priors for the perturber population are summarized in \cref{tab:params-and-priors} and \cref{sec:perturbers}. Our lensing images contain on average about \num{260} \gls*{los} halos and \num{3} subhalos in the lens plane. This means on average about \num{800} parameters are required to describe such images. Likelihood-based sampling of this high-dimensional, transdimensional posterior requires techniques such as reversible-jump \gls*{mcmc} \citep{Daylan:2017kfh,Brewer:2015yya}. To marginalize over the perturber population with \gls*{tmnre}, their parameters are sampled over during data generation but not passed to the ratio estimator.

Since the population of perturbers can contain a member with mass greater than our target subhalo, we need to make this inference task well-defined by ``labeling'' the subhalos of interest. We accomplish this by making the perturber population lighter than the target subhalo, with mass restricted to the range \SIrange{e7}{e8}{\solarmass}. We further assume the target subhalo has been localized to a $\SI{1.4}{''}\times\SI{1.4}{''}$ patch of the image around its true position.

\begin{figure*}
    \centering
    \includegraphics[width=\linewidth]{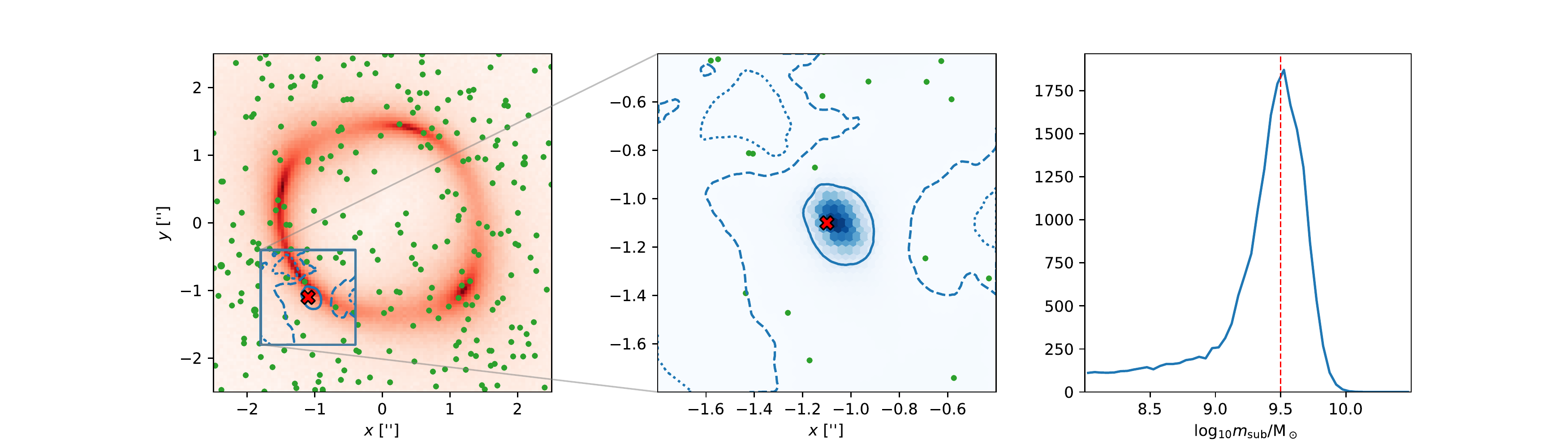}
    \caption{Subhalo position and mass posteriors obtained with \gls*{tmnre}, now marginalizing over a population of \SIrange{e7}{e8}{\solarmass} \gls*{los}/subhalos (green dots) in addition to the unknown macromodel. See the caption of \cref{fig:lss-sub-post} for details. The initial prior for the subhalo's position is indicated by the blue box in the left panel. The range of the $x$-axis in the right panel shows the prior on the $\log_{10}$ of its mass. For both the subhalo's mass and position, the width of the inferred posteriors prevents \gls*{tmnre} from truncating the priors.}
    \label{fig:lshs-post}
\end{figure*}

The final-round inference results for $\psub$ plotted in \cref{fig:lshs-post} show that inclusion of the perturber population has a substantial effect on the posteriors. The posterior for the subhalo's mass peaks around the true value, but has a long tail extending towards the lower boundary of the prior. This indicates we are only able to obtain an upper bound on the subhalo mass rather than a measurement, and cannot exclude the possibility its mass is the lowest value consistent with the prior. Having validated our analysis in simpler cases and checked our ratio estimator has good coverage, we conclude our marginal posteriors are in fact close to the true ones.

Our results are roughly in line with the image segmentation analysis of \citet{Ostdiek:2020cqz,Ostdiek:2020mvo}, which found subhalos of mass above roughly \SI{e8.5}{\solarmass} were resolvable in similar mock observations. In addition, while the 68\% credible region for the subhalo's position contains its true position, the 95\% and 99.7\% credible regions cover nearly the whole prior region.

\begin{figure*}
    \centering
    \includegraphics[width=0.84\linewidth]{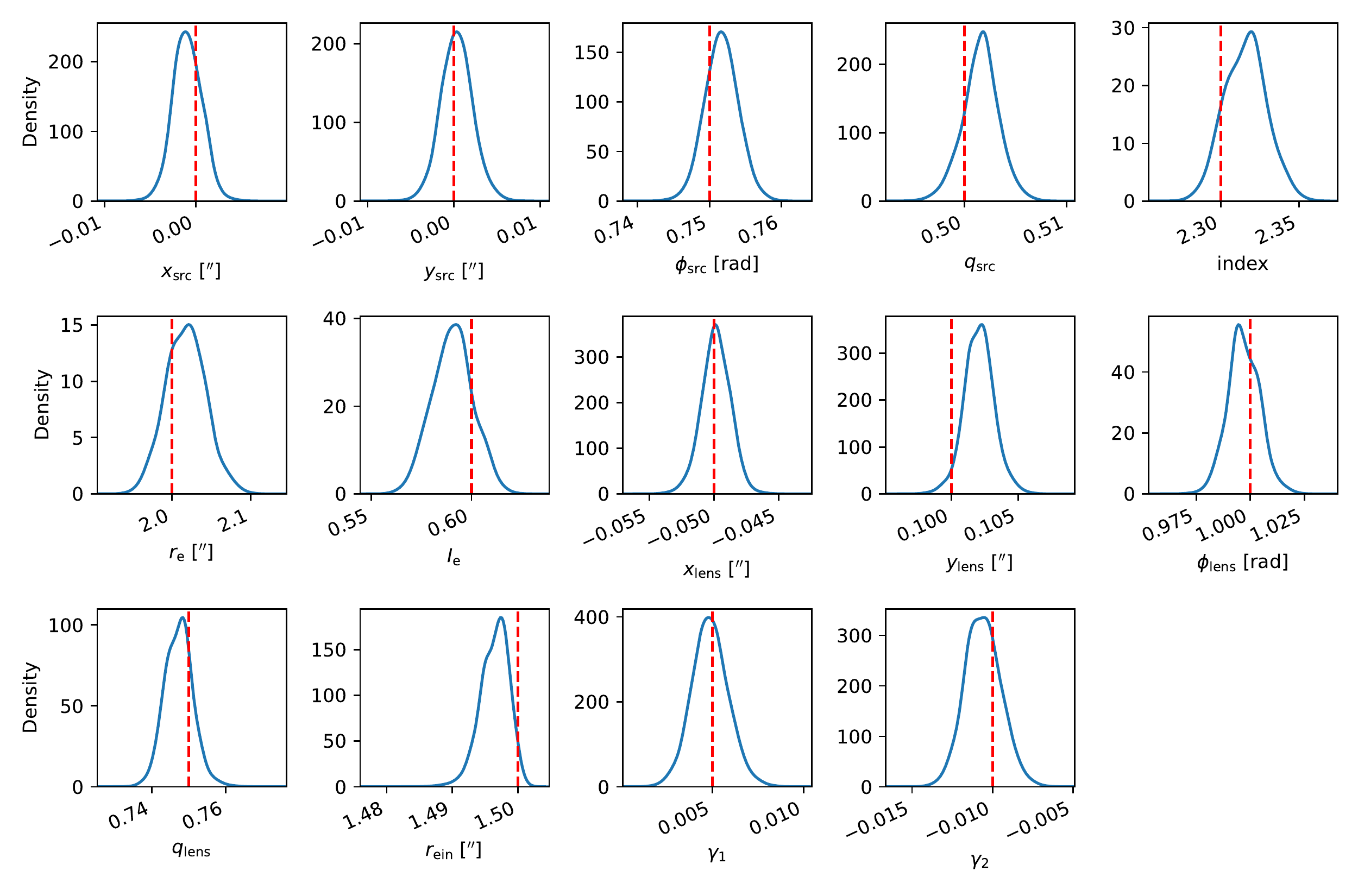}
    \caption{Macromodel 1D marginal posteriors as in \cref{fig:lss-macro-coverage}, but for the inference task where a population of \SIrange{e7}{e8}{\solarmass} are present in the observation. This has the effect of broadening most of the posteriors.}
    \label{fig:lshs-macro-post}
\end{figure*}

The posteriors for the source and lens parameters are shown in \cref{fig:lshs-macro-post}. While some of the parameters' posteriors have comparable widths to those found in the previous inference task (namely $\phi_\mathrm{src/lens}$, $q_\mathrm{src/lens}$, the source index, $I_e$, $\gamma_1$ and $\gamma_2$), others are measured much less precisely due to the stochastic perturber population ($x_\mathrm{src/lens}$, $y_\mathrm{src/lens}$, $r_e$ and $r_\mathrm{Ein}$). We omit coverage plots for this analysis as they are of comparably-good quality to those in the previous subsection.

\section{Discussion and conclusions}

Measuring the properties of individual \gls*{dm} halos on subgalactic scales is an important probe of the fundamental nature of \gls*{dm}. However, extracting their parameters from observations is difficult for a myriad of reasons, including the fact that lenses contain multiple perturbers (sub-/\gls*{los} halos). In this work we demonstrated that \gls*{tmnre} enables analyses of individual perturbers' properties in scenarios where the application of likelihood-based methods is difficult or infeasible. The key strength of \gls*{tmnre} is its ability to directly learn marginal posterior functions for a set of scientifically-interesting parameters from simulated data. By truncating the range of parameters used to generate the simulations, \gls*{tmnre} enables precision inference of individual observations using a targeted set of training data. This enables the previously-intractable marginalization over large perturber populations. Furthermore, the method is applicable to simulators with unknown likelihood functions and large or even variable numbers of input parameters. The resulting inference networks can be poked and prodded to confirm they are statistically well-behaved.

With three lensing simulators of varying complexity, we demonstrated the following characteristics of the method and perturber inference:

\paragraph*{\gls*{tmnre} can recover existing results.} We verified the accuracy of \gls*{tmnre} by confirming it reproduces analytically-calculable posteriors in a toy lensing scenario with known macromodel parameters and subhalo mass.

\paragraph*{\gls*{tmnre} enables statistical checks.} Since the inference networks learned by \gls*{tmnre} are locally amortized over a range of potential observations, we were able to test their statistical consistency. Our checks confirm that \gls*{tmnre} on average produces posteriors with the correct width for the macromodel and subhalo parameters. Such tests would be extremely expensive with likelihood-based inference since they would require rerunning the sampling machinery on numerous mock observations.

\paragraph*{The perturber population matters.} We demonstrated that the sensitivity with which a subhalo's parameters are measurable can be significantly degraded when marginalizing over a population of perturbers. While the $1\sigma$ regions of our position and mass posteriors were centered on the subhalo's true parameters, they had heavy tails extending to the boundaries of our tight, manually-fixed priors. Given our validation, statistical checks and the fact \gls*{tmnre} is so far the only method capable of performing the high-dimensional marginalization required for this analysis, \emph{our results therefore suggest that the population of light perturbers should not be neglected}. It is not, however, excluded that there exists an architecture for the ratio estimator capable of modeling the posterior more accurately than MLP Mixer. This is important to study in future work.

\mbox{}

While this work used simple mock lenses, \gls*{tmnre} makes it possible to add realism and parameters to a simulator without significantly altering the inference procedure, or necessarily increasing the simulation budget \citep{Cole:2021gwr}. It should therefore be straightforward to incorporate various complexities we ignored in this work: a mass-concentration relation for the perturbers, the lens galaxy's light, the (possibly uncertain) \gls*{psf}, multiband observations, drizzling, and even complex noise with an unknown likelihood function. Our analysis can also in principle incorporate more complex source models based on (for example) shapelets~\citep{Birrer:2015rpa,Birrer:2017rpp}, Gaussian processes \citep{Karchev:2021fro} or neural networks \citep{Chianese:2019ifk}. We expect source models capable of refining fine details to improve our measurement precision since the lensing distortions from substructure scale with the gradient of the source.

Another interesting direction for further work is the use of \gls*{tmnre} for model comparison. While here our ratio estimators were trained to compute the likelihood-to-evidence ratio, as pointed out in \citet{aalr} it is possible to learn other ratios of densities. In particular ratio estimators can be used to learn the Bayes factor for assessing the strength of the evidence for different models. This could be used to determine whether an image contains a perturber or not, and to map the minimum-detectable perturber mass as a function of its position.

Overall, we believe using \gls*{tmnre} to measure perturbers as described in this work in combination with measuring the (sub)halo mass function directly \citep{Montel:2022fhv} provides a promising path towards uncovering the identity of dark matter.

\section*{Acknowledgements}

We thank Benjamin Kurt Miller and Elias Dubbeldam for helpful discussions.

A.C. received funding from the Netherlands eScience Center (grant number ETEC.2019.018) and the Schmidt Futures Foundation. This work is part of a project that has received funding from the European Research Council (ERC) under the European Union’s Horizon 2020 research and innovation program (Grant agreement No. 864035).

This work was carried out on the Lisa Compute Cluster at SURFsara, which runs on 100\% wind energy.

Software used: \texttt{astropy} \citep{astropy:2013,astropy:2018}, \texttt{jupyter} \citep{jupyter}, \texttt{matplotlib} \citep{matplotlib}, \texttt{numpy} \citep{numpy}, \texttt{PyTorch} \citep{torch}, \texttt{pytorch-lightning}\footnote{
    \url{https://www.pytorchlightning.ai/}
}, \texttt{seaborn} \citep{seaborn}, \texttt{swyft} \citep{swyftjoss} and \texttt{tqdm} \citep{tqdm}.

\section*{Data Availability}

The data underlying this article will be shared on request to the corresponding author.

\bibliographystyle{mnras}
\bibliography{refs}

\begin{thebibliography}{}
\makeatletter
\relax
\def\mn@urlcharsother{\let\do\@makeother \do\$\do\&\do\#\do\^\do\_\do\%\do\~}
\def\mn@doi{\begingroup\mn@urlcharsother \@ifnextchar [ {\mn@doi@}
  {\mn@doi@[]}}
\def\mn@doi@[#1]#2{\def\@tempa{#1}\ifx\@tempa\@empty \href
  {http://dx.doi.org/#2} {doi:#2}\else \href {http://dx.doi.org/#2} {#1}\fi
  \endgroup}
\def\mn@eprint#1#2{\mn@eprint@#1:#2::\@nil}
\def\mn@eprint@arXiv#1{\href {http://arxiv.org/abs/#1} {{\tt arXiv:#1}}}
\def\mn@eprint@dblp#1{\href {http://dblp.uni-trier.de/rec/bibtex/#1.xml}
  {dblp:#1}}
\def\mn@eprint@#1:#2:#3:#4\@nil{\def\@tempa {#1}\def\@tempb {#2}\def\@tempc
  {#3}\ifx \@tempc \@empty \let \@tempc \@tempb \let \@tempb \@tempa \fi \ifx
  \@tempb \@empty \def\@tempb {arXiv}\fi \@ifundefined
  {mn@eprint@\@tempb}{\@tempb:\@tempc}{\expandafter \expandafter \csname
  mn@eprint@\@tempb\endcsname \expandafter{\@tempc}}}

\bibitem[\protect\citeauthoryear{Alexander, Gleyzer, McDonough, Toomey  \&
  Usai}{Alexander et~al.}{2020}]{Alexander:2019puy}
Alexander S.,  Gleyzer S.,  McDonough E.,  Toomey M.~W.,   Usai E.,  2020,
  \mn@doi [Astrophys. J.] {10.3847/1538-4357/ab7925}, 893, 15

\bibitem[\protect\citeauthoryear{Amorisco et~al.}{Amorisco
  et~al.}{2022}]{Amorisco:2021iim}
Amorisco N.~C.,  et~al., 2022, \mn@doi [Mon. Not. Roy. Astron. Soc.]
  {10.1093/mnras/stab3527}, 510, 2464

\bibitem[\protect\citeauthoryear{Anau~Montel, Coogan, Correa, Karchev  \&
  Weniger}{Anau~Montel et~al.}{2022}]{Montel:2022fhv}
Anau~Montel N.,  Coogan A.,  Correa C.,  Karchev K.,   Weniger C.,  2022

\bibitem[\protect\citeauthoryear{{Astropy Collaboration} et~al.,}{{Astropy
  Collaboration} et~al.}{2013}]{astropy:2013}
{Astropy Collaboration} et~al., 2013, \mn@doi [\aap]
  {10.1051/0004-6361/201322068}, \href
  {http://adsabs.harvard.edu/abs/2013A%26A...558A..33A} {558, A33}

\bibitem[\protect\citeauthoryear{{Astropy Collaboration} et~al.,}{{Astropy
  Collaboration} et~al.}{2018}]{astropy:2018}
{Astropy Collaboration} et~al., 2018, \mn@doi [\aj] {10.3847/1538-3881/aabc4f},
  \href {https://ui.adsabs.harvard.edu/abs/2018AJ....156..123A} {156, 123}

\bibitem[\protect\citeauthoryear{Baltz, Marshall  \& Oguri}{Baltz
  et~al.}{2009}]{Baltz_2009}
Baltz E.~A.,  Marshall P.,   Oguri M.,  2009, \mn@doi [Journal of Cosmology and
  Astrophysics] {10.1088/1475-7516/2009/01/015}, 2009, 015

\bibitem[\protect\citeauthoryear{Bayer, Chatterjee, Koopmans, Vegetti, McKean,
  Treu  \& Fassnacht}{Bayer et~al.}{2018}]{Bayer:2018vhy}
Bayer D.,  Chatterjee S.,  Koopmans L. V.~E.,  Vegetti S.,  McKean J.~P.,  Treu
  T.,   Fassnacht C.~D.,  2018

\bibitem[\protect\citeauthoryear{Birrer, Amara  \& Refregier}{Birrer
  et~al.}{2015}]{Birrer:2015rpa}
Birrer S.,  Amara A.,   Refregier A.,  2015, \mn@doi [Astrophys. J.]
  {10.1088/0004-637X/813/2/102}, 813, 102

\bibitem[\protect\citeauthoryear{Birrer, Amara  \& Refregier}{Birrer
  et~al.}{2017}]{Birrer:2017rpp}
Birrer S.,  Amara A.,   Refregier A.,  2017, \mn@doi [JCAP]
  {10.1088/1475-7516/2017/05/037}, 05, 037

\bibitem[\protect\citeauthoryear{Brehmer, Mishra-Sharma, Hermans, Louppe  \&
  Cranmer}{Brehmer et~al.}{2019}]{Brehmer:2019jyt}
Brehmer J.,  Mishra-Sharma S.,  Hermans J.,  Louppe G.,   Cranmer K.,  2019,
  \mn@doi [Astrophys. J.] {10.3847/1538-4357/ab4c41}, 886, 49

\bibitem[\protect\citeauthoryear{Brewer, Huijser  \& Lewis}{Brewer
  et~al.}{2016}]{Brewer:2015yya}
Brewer B.~J.,  Huijser D.,   Lewis G.~F.,  2016, \mn@doi [Mon. Not. Roy.
  Astron. Soc.] {10.1093/mnras/stv2370}, 455, 1819

\bibitem[\protect\citeauthoryear{Buckley \& Peter}{Buckley \&
  Peter}{2018}]{Buckley:2017ijx}
Buckley M.~R.,  Peter A. H.~G.,  2018, \mn@doi [Phys. Rept.]
  {10.1016/j.physrep.2018.07.003}, 761, 1

\bibitem[\protect\citeauthoryear{Bullock \& Boylan-Kolchin}{Bullock \&
  Boylan-Kolchin}{2017}]{Bullock:2017xww}
Bullock J.~S.,  Boylan-Kolchin M.,  2017, \mn@doi [Ann. Rev. Astron.
  Astrophys.] {10.1146/annurev-astro-091916-055313}, 55, 343

\bibitem[\protect\citeauthoryear{Chianese, Coogan, Hofma, Otten  \&
  Weniger}{Chianese et~al.}{2020}]{Chianese:2019ifk}
Chianese M.,  Coogan A.,  Hofma P.,  Otten S.,   Weniger C.,  2020, \mn@doi
  [Mon. Not. Roy. Astron. Soc.] {10.1093/mnras/staa1477}, 496, 381

\bibitem[\protect\citeauthoryear{Ciotti \& Bertin}{Ciotti \&
  Bertin}{1999}]{Ciotti:1999zs}
Ciotti L.,  Bertin G.,  1999, Astron. Astrophys., 352, 447

\bibitem[\protect\citeauthoryear{Cole, Miller, Witte, Cai, Grootes, Nattino  \&
  Weniger}{Cole et~al.}{2021}]{Cole:2021gwr}
Cole A.,  Miller B.~K.,  Witte S.~J.,  Cai M.~X.,  Grootes M.~W.,  Nattino F.,
   Weniger C.,  2021

\bibitem[\protect\citeauthoryear{Colin, Avila-Reese  \& Valenzuela}{Colin
  et~al.}{2000}]{Colin:2000dn}
Colin P.,  Avila-Reese V.,   Valenzuela O.,  2000, \mn@doi [Astrophys. J.]
  {10.1086/317057}, 542, 622

\bibitem[\protect\citeauthoryear{Collett}{Collett}{2015}]{Collett:2015roa}
Collett T.~E.,  2015, \mn@doi [Astrophys. J.] {10.1088/0004-637X/811/1/20},
  811, 20

\bibitem[\protect\citeauthoryear{{Cranmer}, {Brehmer}  \& {Louppe}}{{Cranmer}
  et~al.}{2020}]{2020PNAS..11730055C}
{Cranmer} K.,  {Brehmer} J.,   {Louppe} G.,  2020, \mn@doi [Proceedings of the
  National Academy of Science] {10.1073/pnas.1912789117}, \href
  {https://ui.adsabs.harvard.edu/abs/2020PNAS..11730055C} {117, 30055}

\bibitem[\protect\citeauthoryear{Dalal \& Kochanek}{Dalal \&
  Kochanek}{2002}]{Dalal:2001fq}
Dalal N.,  Kochanek C.~S.,  2002, \mn@doi [Astrophys. J.] {10.1086/340303},
  572, 25

\bibitem[\protect\citeauthoryear{Daylan, Cyr-Racine, Diaz~Rivero, Dvorkin  \&
  Finkbeiner}{Daylan et~al.}{2018}]{Daylan:2017kfh}
Daylan T.,  Cyr-Racine F.-Y.,  Diaz~Rivero A.,  Dvorkin C.,   Finkbeiner D.~P.,
   2018, \mn@doi [Astrophys. J.] {10.3847/1538-4357/aaaa1e}, 854, 141

\bibitem[\protect\citeauthoryear{Despali \& Vegetti}{Despali \&
  Vegetti}{2017}]{Despali:2016meh}
Despali G.,  Vegetti S.,  2017, \mn@doi [Mon. Not. Roy. Astron. Soc.]
  {10.1093/mnras/stx966}, 469, 1997

\bibitem[\protect\citeauthoryear{Diaz~Rivero, Cyr-Racine  \&
  Dvorkin}{Diaz~Rivero et~al.}{2018}]{DiazRivero:2017xkd}
Diaz~Rivero A.,  Cyr-Racine F.-Y.,   Dvorkin C.,  2018, \mn@doi [Phys. Rev. D]
  {10.1103/PhysRevD.97.023001}, 97, 023001

\bibitem[\protect\citeauthoryear{Dosovitskiy et~al.,}{Dosovitskiy
  et~al.}{2020}]{vit}
Dosovitskiy A.,  et~al., 2020, CoRR, abs/2010.11929

\bibitem[\protect\citeauthoryear{Efstathiou}{Efstathiou}{1992}]{Efstathiou:1992zz}
Efstathiou G.,  1992, Mon. Not. Roy. Astron. Soc., 256, 43P

\bibitem[\protect\citeauthoryear{Fitts et~al.}{Fitts
  et~al.}{2017}]{Fitts:2016usl}
Fitts A.,  et~al., 2017, \mn@doi [Mon. Not. Roy. Astron. Soc.]
  {10.1093/mnras/stx1757}, 471, 3547

\bibitem[\protect\citeauthoryear{Gilman, Birrer, Nierenberg, Treu, Du  \&
  Benson}{Gilman et~al.}{2020}]{Gilman:2019nap}
Gilman D.,  Birrer S.,  Nierenberg A.,  Treu T.,  Du X.,   Benson A.,  2020,
  \mn@doi [Mon. Not. Roy. Astron. Soc.] {10.1093/mnras/stz3480}, 491, 6077

\bibitem[\protect\citeauthoryear{{Giocoli}, {Tormen}, {Sheth}  \& {van den
  Bosch}}{{Giocoli} et~al.}{2010}]{2010MNRAS.404..502G}
{Giocoli} C.,  {Tormen} G.,  {Sheth} R.~K.,   {van den Bosch} F.~C.,  2010,
  \mn@doi [\mnras] {10.1111/j.1365-2966.2010.16311.x}, \href
  {https://ui.adsabs.harvard.edu/abs/2010MNRAS.404..502G} {404, 502}

\bibitem[\protect\citeauthoryear{Greenberg, Nonnenmacher  \& Macke}{Greenberg
  et~al.}{2019}]{DBLP:journals/corr/abs-1905-07488}
Greenberg D.~S.,  Nonnenmacher M.,   Macke J.~H.,  2019, CoRR, abs/1905.07488

\bibitem[\protect\citeauthoryear{Gu et~al.}{Gu et~al.}{2022}]{Gu:2022xhk}
Gu A.,  et~al., 2022

\bibitem[\protect\citeauthoryear{Harris et~al.,}{Harris et~al.}{2020}]{numpy}
Harris C.~R.,  et~al., 2020, \mn@doi [Nature] {10.1038/s41586-020-2649-2}, 585,
  357

\bibitem[\protect\citeauthoryear{{He} et~al.,}{{He}
  et~al.}{2020}]{2020arXiv201013221H}
{He} Q.,  et~al., 2020, arXiv e-prints, \href
  {https://ui.adsabs.harvard.edu/abs/2020arXiv201013221H} {p. arXiv:2010.13221}

\bibitem[\protect\citeauthoryear{{Hermans}, {Begy}  \& {Louppe}}{{Hermans}
  et~al.}{2019}]{aalr}
{Hermans} J.,  {Begy} V.,   {Louppe} G.,  2019, arXiv e-prints, \href
  {https://ui.adsabs.harvard.edu/abs/2019arXiv190304057H} {p. arXiv:1903.04057}

\bibitem[\protect\citeauthoryear{{Hermans}, {Delaunoy}, {Rozet}, {Wehenkel}  \&
  {Louppe}}{{Hermans} et~al.}{2021}]{averting}
{Hermans} J.,  {Delaunoy} A.,  {Rozet} F.,  {Wehenkel} A.,   {Louppe} G.,
  2021, arXiv e-prints, \href
  {https://ui.adsabs.harvard.edu/abs/2021arXiv211006581H} {p. arXiv:2110.06581}

\bibitem[\protect\citeauthoryear{Hezaveh, Dalal, Holder, Kisner, Kuhlen  \&
  Perreault~Levasseur}{Hezaveh et~al.}{2016a}]{Hezaveh:2014aoa}
Hezaveh Y.,  Dalal N.,  Holder G.,  Kisner T.,  Kuhlen M.,
  Perreault~Levasseur L.,  2016a, \mn@doi [JCAP]
  {10.1088/1475-7516/2016/11/048}, 11, 048

\bibitem[\protect\citeauthoryear{Hezaveh et~al.}{Hezaveh
  et~al.}{2016b}]{Hezaveh:2016ltk}
Hezaveh Y.~D.,  et~al., 2016b, \mn@doi [Astrophys. J.]
  {10.3847/0004-637X/823/1/37}, 823, 37

\bibitem[\protect\citeauthoryear{Hogan \& Dalcanton}{Hogan \&
  Dalcanton}{2000}]{Hogan:2000bv}
Hogan C.~J.,  Dalcanton J.~J.,  2000, \mn@doi [Phys. Rev. D]
  {10.1103/PhysRevD.62.063511}, 62, 063511

\bibitem[\protect\citeauthoryear{Hu, Barkana  \& Gruzinov}{Hu
  et~al.}{2000}]{Hu:2000ke}
Hu W.,  Barkana R.,   Gruzinov A.,  2000, \mn@doi [Phys. Rev. Lett.]
  {10.1103/PhysRevLett.85.1158}, 85, 1158

\bibitem[\protect\citeauthoryear{Hunter}{Hunter}{2007}]{matplotlib}
Hunter J.~D.,  2007, \mn@doi [Computing in Science \& Engineering]
  {10.1109/MCSE.2007.55}, 9, 90

\bibitem[\protect\citeauthoryear{Karchev, Coogan  \& Weniger}{Karchev
  et~al.}{2022}]{Karchev:2021fro}
Karchev K.,  Coogan A.,   Weniger C.,  2022, \mn@doi [Mon. Not. Roy. Astron.
  Soc.] {10.1093/mnras/stac311}, 512, 661

\bibitem[\protect\citeauthoryear{Kluyver et~al.,}{Kluyver
  et~al.}{2016}]{jupyter}
Kluyver T.,  et~al., 2016, in Loizides F.,  Scmidt B.,  eds, Positioning and
  Power in Academic Publishing: Players, Agents and Agendas. IOS Press, pp
  87--90, \url {https://eprints.soton.ac.uk/403913/}

\bibitem[\protect\citeauthoryear{Klypin, Kravtsov, Valenzuela  \& Prada}{Klypin
  et~al.}{1999}]{Klypin:1999uc}
Klypin A.~A.,  Kravtsov A.~V.,  Valenzuela O.,   Prada F.,  1999, \mn@doi
  [Astrophys. J.] {10.1086/307643}, 522, 82

\bibitem[\protect\citeauthoryear{Koopmans}{Koopmans}{2006}]{Koopmans:2005ig}
Koopmans L. V.~E.,  2006, \mn@doi [EAS Publ. Ser.] {10.1051/eas:2006064}, 20,
  161

\bibitem[\protect\citeauthoryear{{Lueckmann}, {Goncalves}, {Bassetto},
  {{\"O}cal}, {Nonnenmacher}  \& {Macke}}{{Lueckmann}
  et~al.}{2017}]{2017arXiv171101861L}
{Lueckmann} J.-M.,  {Goncalves} P.~J.,  {Bassetto} G.,  {{\"O}cal} K.,
  {Nonnenmacher} M.,   {Macke} J.~H.,  2017, arXiv e-prints, \href
  {https://ui.adsabs.harvard.edu/abs/2017arXiv171101861L} {p. arXiv:1711.01861}

\bibitem[\protect\citeauthoryear{Mao \& Schneider}{Mao \&
  Schneider}{1998}]{Mao:1997ek}
Mao S.-d.,  Schneider P.,  1998, \mn@doi [Mon. Not. Roy. Astron. Soc.]
  {10.1046/j.1365-8711.1998.01319.x}, 295, 587

\bibitem[\protect\citeauthoryear{Meneghetti}{Meneghetti}{2016}]{lensing}
Meneghetti M.,  2016, Introduction to Gravitational Lensing

\bibitem[\protect\citeauthoryear{Miller, Cole, Louppe  \& Weniger}{Miller
  et~al.}{2020}]{swyft}
Miller B.~K.,  Cole A.,  Louppe G.,   Weniger C.,  2020

\bibitem[\protect\citeauthoryear{Miller, Cole, Forr\'{e}, Louppe  \&
  Weniger}{Miller et~al.}{2021}]{tmnre}
Miller B.~K.,  Cole A.,  Forr\'{e} P.,  Louppe G.,   Weniger C.,  2021, in
  Ranzato M.,  Beygelzimer A.,  Dauphin Y.,  Liang P.,   Vaughan J.~W.,  eds, ~
  Vol. 34, Advances in Neural Information Processing Systems. Curran
  Associates, Inc., pp 129--143, \url
  {https://proceedings.neurips.cc/paper/2021/file/01632f7b7a127233fa1188bd6c2e42e1-Paper.pdf}

\bibitem[\protect\citeauthoryear{Miller, Cole, Weniger, Nattino, Ku  \&
  Grootes}{Miller et~al.}{2022}]{swyftjoss}
Miller B.~K.,  Cole A.,  Weniger C.,  Nattino F.,  Ku O.,   Grootes M.~W.,
  2022, \mn@doi [Journal of Open Source Software] {10.21105/joss.04205}, 7,
  4205

\bibitem[\protect\citeauthoryear{Moore, Ghigna, Governato, Lake, Quinn, Stadel
  \& Tozzi}{Moore et~al.}{1999}]{Moore:1999nt}
Moore B.,  Ghigna S.,  Governato F.,  Lake G.,  Quinn T.~R.,  Stadel J.,
  Tozzi P.,  1999, \mn@doi [Astrophys. J. Lett.] {10.1086/312287}, 524, L19

\bibitem[\protect\citeauthoryear{{Nightingale} \& {Dye}}{{Nightingale} \&
  {Dye}}{2015}]{2015MNRAS.452.2940N}
{Nightingale} J.~W.,  {Dye} S.,  2015, \mn@doi [\mnras]
  {10.1093/mnras/stv1455}, \href
  {https://ui.adsabs.harvard.edu/abs/2015MNRAS.452.2940N} {452, 2940}

\bibitem[\protect\citeauthoryear{{O'Riordan}, {Warren}  \&
  {Mortlock}}{{O'Riordan} et~al.}{2020}]{2020MNRAS.496.3424O}
{O'Riordan} C.~M.,  {Warren} S.~J.,   {Mortlock} D.~J.,  2020, \mn@doi [\mnras]
  {10.1093/mnras/staa1697}, \href
  {https://ui.adsabs.harvard.edu/abs/2020MNRAS.496.3424O} {496, 3424}

\bibitem[\protect\citeauthoryear{Ostdiek, Diaz~Rivero  \& Dvorkin}{Ostdiek
  et~al.}{2022a}]{Ostdiek:2020cqz}
Ostdiek B.,  Diaz~Rivero A.,   Dvorkin C.,  2022a, \mn@doi [Astron. Astrophys.]
  {10.1051/0004-6361/202142030}, 657, L14

\bibitem[\protect\citeauthoryear{Ostdiek, Diaz~Rivero  \& Dvorkin}{Ostdiek
  et~al.}{2022b}]{Ostdiek:2020mvo}
Ostdiek B.,  Diaz~Rivero A.,   Dvorkin C.,  2022b, \mn@doi [Astrophys. J.]
  {10.3847/1538-4357/ac2d8d}, 927, 83

\bibitem[\protect\citeauthoryear{{Papamakarios} \& {Murray}}{{Papamakarios} \&
  {Murray}}{2016}]{2016arXiv160506376P}
{Papamakarios} G.,  {Murray} I.,  2016, arXiv e-prints, \href
  {https://ui.adsabs.harvard.edu/abs/2016arXiv160506376P} {p. arXiv:1605.06376}

\bibitem[\protect\citeauthoryear{{Papamakarios}, {Sterratt}  \&
  {Murray}}{{Papamakarios} et~al.}{2018}]{2018arXiv180507226P}
{Papamakarios} G.,  {Sterratt} D.~C.,   {Murray} I.,  2018, arXiv e-prints,
  \href {https://ui.adsabs.harvard.edu/abs/2018arXiv180507226P} {p.
  arXiv:1805.07226}

\bibitem[\protect\citeauthoryear{Paszke et~al.,}{Paszke et~al.}{2019}]{torch}
Paszke A.,  et~al., 2019, in Wallach H.,  Larochelle H.,  Beygelzimer A.,
  d\textquotesingle Alch\'{e}-Buc F.,  Fox E.,   Garnett R.,  eds, , Advances
  in Neural Information Processing Systems 32.
Curran Associates, Inc., pp 8024--8035, \url
  {http://papers.neurips.cc/paper/9015-pytorch-an-imperative-style-high-performance-deep-learning-library.pdf}

\bibitem[\protect\citeauthoryear{{Planck Collaboration}}{{Planck
  Collaboration}}{2020}]{Planck2018}
{Planck Collaboration} 2020, \mn@doi [Astronomy and Astrophysics]
  {10.1051/0004-6361/201833910}, 641, A6

\bibitem[\protect\citeauthoryear{Profumo}{Profumo}{2017}]{Profumo:2017hqp}
Profumo S.,  2017, {An Introduction to Particle Dark Matter}.
World Scientific, \mn@doi{10.1142/q0001}

\bibitem[\protect\citeauthoryear{Richings, Frenk, Jenkins, Robertson  \&
  Schaller}{Richings et~al.}{2021}]{Richings:2020auv}
Richings J.,  Frenk C.,  Jenkins A.,  Robertson A.,   Schaller M.,  2021,
  \mn@doi [Mon. Not. Roy. Astron. Soc.] {10.1093/mnras/staa4013}, 501, 4657

\bibitem[\protect\citeauthoryear{Schneider \& Sluse}{Schneider \&
  Sluse}{2013}]{Schneider:2013sxa}
Schneider P.,  Sluse D.,  2013, \mn@doi [Astron. Astrophys.]
  {10.1051/0004-6361/201321882}, 559, A37

\bibitem[\protect\citeauthoryear{Schneider \& Sluse}{Schneider \&
  Sluse}{2014}]{Schneider:2013wga}
Schneider P.,  Sluse D.,  2014, \mn@doi [Astron. Astrophys.]
  {10.1051/0004-6361/201322106}, 564, A103

\bibitem[\protect\citeauthoryear{{Skilling}}{{Skilling}}{2004}]{2004AIPC..735..395S}
{Skilling} J.,  2004, in {Fischer} R.,  {Preuss} R.,   {Toussaint} U.~V.,  eds,
   American Institute of Physics Conference Series Vol. 735, Bayesian Inference
  and Maximum Entropy Methods in Science and Engineering: 24th International
  Workshop on Bayesian Inference and Maximum Entropy Methods in Science and
  Engineering. pp 395--405, \mn@doi{10.1063/1.1835238}

\bibitem[\protect\citeauthoryear{Spergel \& Steinhardt}{Spergel \&
  Steinhardt}{2000}]{Spergel:1999mh}
Spergel D.~N.,  Steinhardt P.~J.,  2000, \mn@doi [Phys. Rev. Lett.]
  {10.1103/PhysRevLett.84.3760}, 84, 3760

\bibitem[\protect\citeauthoryear{Suyu, Marshall, Blandford, Fassnacht,
  Koopmans, McKean  \& Treu}{Suyu et~al.}{2009}]{Suyu:2008zp}
Suyu S.~H.,  Marshall P.~J.,  Blandford R.~D.,  Fassnacht C.~D.,  Koopmans L.
  V.~E.,  McKean J.~P.,   Treu T.,  2009, \mn@doi [Astrophys. J.]
  {10.1088/0004-637X/691/1/277}, 691, 277

\bibitem[\protect\citeauthoryear{Tessore \& Metcalf}{Tessore \&
  Metcalf}{2015}]{Tessore:2015baa}
Tessore N.,  Metcalf R.~B.,  2015, \mn@doi [Astron. Astrophys.]
  {10.1051/0004-6361/201526773}, 580, A79

\bibitem[\protect\citeauthoryear{Tinker, Kravtsov, Klypin, Abazajian, Warren,
  Yepes, Gottlober  \& Holz}{Tinker et~al.}{2008}]{Tinker:2008ff}
Tinker J.~L.,  Kravtsov A.~V.,  Klypin A.,  Abazajian K.,  Warren M.~S.,  Yepes
  G.,  Gottlober S.,   Holz D.~E.,  2008, \mn@doi [Astrophys. J.]
  {10.1086/591439}, 688, 709

\bibitem[\protect\citeauthoryear{Tolstikhin et~al.,}{Tolstikhin
  et~al.}{2021}]{mlpmixer}
Tolstikhin I.~O.,  et~al., 2021, CoRR, abs/2105.01601

\bibitem[\protect\citeauthoryear{Vegetti \& Koopmans}{Vegetti \&
  Koopmans}{2009a}]{Vegetti:2008eg}
Vegetti S.,  Koopmans L. V.~E.,  2009a, \mn@doi [Mon. Not. Roy. Astron. Soc.]
  {10.1111/j.1365-2966.2008.14005.x}, 392, 945

\bibitem[\protect\citeauthoryear{{Vegetti} \& {Koopmans}}{{Vegetti} \&
  {Koopmans}}{2009b}]{2009MNRAS.400.1583V}
{Vegetti} S.,  {Koopmans} L.~V.~E.,  2009b, \mn@doi [\mnras]
  {10.1111/j.1365-2966.2009.15559.x}, \href
  {https://ui.adsabs.harvard.edu/abs/2009MNRAS.400.1583V} {400, 1583}

\bibitem[\protect\citeauthoryear{{Vegetti}, {Czoske}  \& {Koopmans}}{{Vegetti}
  et~al.}{2010a}]{2010MNRAS.407..225V}
{Vegetti} S.,  {Czoske} O.,   {Koopmans} L. V.~E.,  2010a, \mn@doi [\mnras]
  {10.1111/j.1365-2966.2010.16952.x}, \href
  {https://ui.adsabs.harvard.edu/abs/2010MNRAS.407..225V} {407, 225}

\bibitem[\protect\citeauthoryear{{Vegetti}, {Koopmans}, {Bolton}, {Treu}  \&
  {Gavazzi}}{{Vegetti} et~al.}{2010b}]{2010MNRAS.408.1969V}
{Vegetti} S.,  {Koopmans} L.~V.~E.,  {Bolton} A.,  {Treu} T.,   {Gavazzi} R.,
  2010b, \mn@doi [\mnras] {10.1111/j.1365-2966.2010.16865.x}, \href
  {https://ui.adsabs.harvard.edu/abs/2010MNRAS.408.1969V} {408, 1969}

\bibitem[\protect\citeauthoryear{{Vegetti}, {Lagattuta}, {McKean}, {Auger},
  {Fassnacht}  \& {Koopmans}}{{Vegetti} et~al.}{2012}]{2012Natur.481..341V}
{Vegetti} S.,  {Lagattuta} D.~J.,  {McKean} J.~P.,  {Auger} M.~W.,  {Fassnacht}
  C.~D.,   {Koopmans} L.~V.~E.,  2012, \mn@doi [\nat] {10.1038/nature10669},
  \href {https://ui.adsabs.harvard.edu/abs/2012Natur.481..341V} {481, 341}

\bibitem[\protect\citeauthoryear{Wagner-Carena, Aalbers, Birrer, Nadler,
  Darragh-Ford, Marshall  \& Wechsler}{Wagner-Carena
  et~al.}{2022}]{Wagner-Carena:2022mrn}
Wagner-Carena S.,  Aalbers J.,  Birrer S.,  Nadler E.~O.,  Darragh-Ford E.,
  Marshall P.~J.,   Wechsler R.~H.,  2022

\bibitem[\protect\citeauthoryear{Waskom}{Waskom}{2021}]{seaborn}
Waskom M.~L.,  2021, \mn@doi [Journal of Open Source Software]
  {10.21105/joss.03021}, 6, 3021

\bibitem[\protect\citeauthoryear{Zhang, Mishra-Sharma  \& Dvorkin}{Zhang
  et~al.}{2022}]{Zhang:2022djp}
Zhang G.,  Mishra-Sharma S.,   Dvorkin C.,  2022

\bibitem[\protect\citeauthoryear{\c{C}aǧan \c{S}eng\"ul, Tsang, Diaz~Rivero,
  Dvorkin, Zhu  \& Seljak}{\c{C}aǧan \c{S}eng\"ul
  et~al.}{2020}]{CaganSengul:2020nat}
\c{C}aǧan \c{S}eng\"ul A.,  Tsang A.,  Diaz~Rivero A.,  Dvorkin C.,  Zhu
  H.-M.,   Seljak U.,  2020, \mn@doi [Phys. Rev. D]
  {10.1103/PhysRevD.102.063502}, 102, 063502

\bibitem[\protect\citeauthoryear{da Costa-Luis et~al.,}{da~Costa-Luis
  et~al.}{2022}]{tqdm}
da Costa-Luis C.,  et~al., 2022, {tqdm: A fast, Extensible Progress Bar for
  Python and CLI}, \mn@doi{10.5281/zenodo.6412640}, \url
  {https://doi.org/10.5281/zenodo.6412640}

\bibitem[\protect\citeauthoryear{de Blok \& McGaugh}{de~Blok \&
  McGaugh}{1997}]{deBlok:1997zlw}
de Blok W. J.~G.,  McGaugh S.~S.,  1997, \mn@doi [Mon. Not. Roy. Astron. Soc.]
  {10.1093/mnras/290.3.533}, 290, 533

\makeatother
\end{thebibliography}

\appendix

\section{Compression network architectures}
\label{app:architectures}

The compressor architectures are given in \cref{tab:cnn} and \cref{tab:mixer}. Note that we standardize the images before providing them to the networks.

\begin{table}
    \caption{The convolutional compression network used in the macromodel parameter ratio estimator. The notation is taken from \texttt{PyTorch}: the arguments to \texttt{Conv2d} are the number of input channels, output channels, kernel size, stride and padding, respectively. The horizontal lines demarcate where the number of channels changes. The output of the network is flattened into a vector with \num{128} features.}
    \centering
    \begin{tabular}{c}
        \toprule
        \texttt{Conv2d(1, 4, 8, 2, 1, bias=False)} \\
        \texttt{BatchNorm2d(4)} \\
        \texttt{LeakyReLU(0.2)} \\
        \midrule
        \texttt{Conv2d(4, 8, 8, 2, 1, bias=False)} \\
        \texttt{BatchNorm2d(8)} \\
        \texttt{LeakyReLU(0.2)} \\
        \midrule
        \texttt{Conv2d(8, 16, 8, 2, 1, bias=False)} \\
        \texttt{BatchNorm2d(16)} \\
        \texttt{LeakyReLU(0.2)} \\
        \midrule
        \texttt{Conv2d(16, 32, 8, 2, 1, bias=False)} \\
        \texttt{BatchNorm2d(32)} \\
        \texttt{LeakyReLU(0.2)} \\
        \bottomrule
    \end{tabular}
    \label{tab:cnn}
\end{table}

\begin{table}
    \centering
    \caption{The details of the MLP Mixer compression network in the subhalo ratio estimator. We use the implementation from \url{https://github.com/lucidrains/mlp-mixer-pytorch}, with arguments given in the table.}
    \begin{tabular}{r l}
        \toprule
        \texttt{image\_size} & \texttt{100} \\
        \texttt{channels} & \texttt{1} \\
        \texttt{patch\_size} & \texttt{10}\\
        \texttt{dim} & \texttt{256} \\
        \texttt{depth} & \texttt{4} \\
        \texttt{num\_classes} & 32 \\
        \texttt{dropout} & \texttt{0.1} \\
        \bottomrule
    \end{tabular}
    \label{tab:mixer}
\end{table}

\bsp	
\label{lastpage}
\end{document}